\documentclass[aps,amsmath,twocolumn,amssymb,floatfix,showpacs,superscriptaddress,nofootinbib,longbibliography]{revtex4-1}
\usepackage{mathtools}
\usepackage{braket}
\usepackage[dvipsnames]{xcolor}
\usepackage{float}
\usepackage{subfigure}
\usepackage{dsfont}
\usepackage[colorlinks=true,linktoc=page,linkcolor=PineGreen,citecolor=purple,urlcolor=violet]{hyperref}
\newcommand{\vect}[1]{\boldsymbol{\mathrm{#1}}}
\mathchardef\mhyphen="2D % Define a "math hyphen"

\newcommand{\ie}{{i.e.,\,\,}}

\newcommand\bea{\begin{eqnarray}}
\newcommand\eea{\end{eqnarray}}
\newcommand\beq{\begin{equation}}  
\newcommand\eeq{\end{equation}}

\newcommand{\non}{\nonumber}

\usepackage[normalem]{ulem}
\definecolor{lime}{HTML}{A6CE39}
\usepackage{sidecap,tikz}
\DeclareRobustCommand{\orcidicon}{\hspace{-1.0mm}
	\begin{tikzpicture}
		\draw[lime, fill=lime] (0.0,0.0) 
		circle [radius=0.15] 
		node[white] {{\fontfamily{qag}\selectfont \tiny \,ID}};
		\draw[white, fill=white] (-0.0525,0.095) 
		circle [radius=0.007];
	\end{tikzpicture}
	\hspace{-3.0mm}
}
\foreach \x in {A, ..., Z}{\expandafter\xdef\csname orcid\x\endcsname{\noexpand\href{https://orcid.org/\csname orcidauthor\x\endcsname}
		{\noexpand\orcidicon}}
}

\AtBeginDocument{%
	\newwrite\bibnotes
	\def\bibnotesext{Notes.bib}
	\immediate\openout\bibnotes=\jobname\bibnotesext
	\immediate\write\bibnotes{@CONTROL{REVTEX41Control}}
	\immediate\write\bibnotes{@CONTROL{%
			apsrev41Control,author="08",editor="1",pages="1",title="1",year="1"}}
	\if@filesw
	\immediate\write\@auxout{\string\citation{apsrev41Control}}%
	\fi
}%
\begin{document}

%=============START of MAIN PAPER===============

\title{Corner modes in non-Hermitian next-nearest-neighbor hopping model}

\author{Arnob Kumar Ghosh}
\affiliation{Institute of Physics, Sachivalaya Marg, Bhubaneswar-751005, India}
\affiliation{Homi Bhabha National Institute, Training School Complex, Anushakti Nagar, Mumbai 400094, India}
\affiliation{Department of Physics and Astronomy, Uppsala University, Box 516, 75120 Uppsala, Sweden}

\author{Arijit Saha\orcidC{}}
\email{arijit@iopb.res.in}
\affiliation{Institute of Physics, Sachivalaya Marg, Bhubaneswar-751005, India}
\affiliation{Homi Bhabha National Institute, Training School Complex, Anushakti Nagar, Mumbai 400094, India}

\author{Tanay Nag\orcidB{}}
\email{tanay.nag@hyderabad.bits-pilani.ac.in}
\affiliation{Department of Physics, BITS Pilani-Hyderabad Campus, Telangana 500078, India}

%--------------------------------------------------------
%--------------------------------------------------------
\begin{abstract}
We consider a non-Hermitian (NH) analog of a second-order topological insulator, protected by chiral symmetry, in the presence of next-nearest neighbor hopping elements to theoretically investigate 
the interplay beyond the first nearest neighbor hopping amplitudes and topological order away from Hermiticity. In addition to the four zero-energy corner modes present in the first nearest neighbor hopping model, we uncover that the second nearest neighbor hopping introduces another topological phase with sixteen zero-energy corner modes. Importantly, the NH effects are manifested in altering the Hermitian phase boundaries for both the models. While comparing the complex energy spectrum under open boundary conditions, and bi-orthogonalized quadrupolar winding number in real space, we resolve the apparent anomaly in the bulk boundary correspondence of the NH system as compared to the Hermitian counterpart by incorporating the effect of non-Bloch form 
of momentum into the mass term. The above invariant is also capable of capturing the phase boundaries between the two different topological phases where the degeneracy of the corner modes is evident, as exclusively observed for the second nearest neighbor model. 
\end{abstract}
%--------------------------------------------------------
%--------------------------------------------------------

\maketitle

%======================================================
\section{Introduction}
\label{intro}
%======================================================

The systems with topological band properties are identified with gapless boundary modes that are characterized by symmetry-protected topological invariants. This is known as bulk boundary correspondence (BBC)~\cite{hasan2010colloquium,sato2017topological}. The conventional BBC, 
being an integral part of the first-order ($n=1$) topological phase~\cite{hasan2010colloquium,Haldane88,Kane05,bernevig2006quantum}, is generalized for higher-order ($n>1$) topological phases in 
$d\ge2$ dimensions where there exist $n_c=(d-n)$-dimensional boundary modes~\cite{benalcazar2017,benalcazarprb2017,Song2017,Langbehn2017,schindler2018,Franca2018,wang2018higher,Roy2019,Szumniak2020,Ni2020,BiyeXie2021,saha2022dipolar,Zhu2018,Liu2018,Yan2018,WangWeak2018,Zhang2019,ZhangFe2019PRL,Volpez2019,YanPRB2019,Ghorashi2019,Wu2020,jelena2020HOTSC,BitanTSC2020,SongboPRR22020,kheirkhah2020vortex,YanPRL2019,AhnPRL2020,luo2021higherorder2021,QWang2018,Ghosh2021PRB,RoyPRBL2021,li2021higher,ghosh2022systematic,ghosh2022dynamical}. For example, the second-order topological insulator (SOTI) in two dimensions hosts zero-dimensional (0D) localized corner modes at zero-energy while this bulk phase is characterized by nested polarization or quadrupolar moment. Very recently, it has been reported that the number of boundary modes in a topological phase can be tuned by considering next-nearest-neighbor hopping terms~\cite{DeGottardi13,Niu12,xie2019topological,Dias22,mondal2024persistent} 
as well as by implementing periodic Floquet drive~\cite{Thakurathi2013,ghosh2022systematic}.  
Once the extended model continues to preserve the chiral symmetry (CS), one can characterize the 
new topological phase by winding numbers in odd spatial dimension~ \cite{Schnyder08,kitaev2009periodic,ryu2010topological}. The number of degenerate zero-energy states at each boundary increases according to the enhancement of the range of the hopping amplitudes as indicated by the winding number ensuring the BBC~\cite{Ryu02,Teo10}. It is noteworthy that the one-dimensional winding number for the first-order topological systems becomes passive in the case of even-dimensional generalizations~\cite{nakahara2018geometry}.
%even-dimensional generalizations of the one-dimensional winding number for the first-order topological systems do not work out  \cite{nakahara2018geometry}. 
In contrast, the higher-order topological (HOT) phase in even spatial dimension can be characterized by an appropriately defined winding number preserving CS as a constraint~\cite{Benalcazar22}. 

In recent years, thanks to the practical realization of higher-order topological phases in meta-materials~\cite{parappurath2020direct,yang2019realization,Malzard15,regensburger2012parity} where energy conservation no longer holds~\cite{el2018non,Denner2021} and the domain of topological quantum matter can be extended to the non-Hermitian (NH) systems. The coupling to the environment~ \cite{Bergholtz19,Yang21,San-Jose2016}, disorder/interaction-mediated quasiparticles with finite lifetime~\cite{kozii2017non,Yoshida18,Shen18} can effectively induce complex self energy that is modeled by an NH effective Hamiltonian~ \cite{Musslimani08,Makris08,YaoPRLSecond2018,KawabataPRX2019,Bergholtz2021,Sone2020,
Denner2021,ghosh2024majorana}. Interestingly, the non-Bloch nature of the wavefunction for the NH systems renormalizes the topological mass term, thus enriching the BBC such that topological phase transitions perceived with open-boundary conditions can be explained by an appropriate bulk invariant~\cite{YaoPRL2018,Kunst18,NoriNHPRL2019,helbig2020generalized,Borgnia20,Koch2020,
ZirnsteinPRL2021,TakaneJPS2021}. The NH topological systems showcase various intriguing features such as the skin effect where the bulk states accumulate at the boundary~ \cite{YaoPRL2018,Kunst18,helbig2020generalized,Kawabata20b}, exceptional points where eigenstates, corresponding to the degenerate bands, coalesce~\cite{bender2007making,heiss2012physics}. 

Going beyond the scope of the first nearest neighbor (NN) hopping, the second NN or the next-nearest-neighbor hopping elements in Hermitian systems is found to mediate versatile topological phases where the number of zero-energy modes increases~\cite{Beatriz19,Dias22,Fangzhao18,Hsu20}. In this context, the interplay between the next-NN hopping elements and non-Hermiticity is still in its infancy as far as HOT systems are concerned~\cite{NoriNHPRL2019,Ghosh22NH,Liu21,Liu23,RangiPRB2024,JiPRB2024}. Therefore, considering a two-dimensional (2D) NH SOTI with next-NN hopping, we, therefore, examine whether non-Hermiticity induces exceptional SOTI phases, otherwise absent in the Hermitian case, and address the following interesting questions that have not been explored so far in the literature, to the best of our knowledge: How does the BBC change in the above NH phases?  Can we characterize these emerging exceptional topological phases by bi-orthogonalized non-Bloch winding number? 

We consider a CS-preserved generic model, hosting SOTI phases in the presence of non-Hermiticity and next-NN hopping terms. The first [second] NN Hermitian model Hamiltonian can host four [four and sixteen] zero-energy corner modes while exceptional points, caused by the NH effects, reshape topological phase boundaries as compared to the Hermitian case. As a result, 
we find that BBC is not only different from their Hermitian counterparts but also non-trivially modified once second NN hopping terms are included. The phase boundaries between four [one] and one [none] zero-energy modes per corner in the second [first] NN model are revised following the dressed mass term due to the non-Bloch nature of wavefunction for NH Hamiltonian. We compute the quadrupolar winding number (QWN) appropriately in real space by exploiting the CS as well as implementing the bi-orthogonalization and non-Bloch nature to lay out the phase diagram, which is in accordance with the complex spectrum under open boundary conditions (OBC). 

The remainder of the article is organized in the following way. We discuss the details of the tight-binding Hamiltonian in Sec.~\ref{model}, where both the NH first and second hopping models are demonstrated. Sec.~\ref{result} is devoted to the main results of this article. In particular, we discuss the result associated with the first and the second NN hopping models in Secs.~\ref{firstNN} and \ref{secondNN}, respectively. Next, we illustrate the exceptional phase diagram in Sec.~\ref{qwn} by examining the QWN. We finally summarize and conclude in Sec.~\ref{conclusion}. In Appendices~\ref{App:A} and \ref{App:B}, we discuss the spatial symmetries of our model, the effect of asymmetric hoppings, and the finite size effect with the NH second NN hopping model.

%======================================================
\section{Model Hamiltonian}\label{model}
%======================================================

We consider the SOTI model in the presence of the second NN hopping as follows~\cite{Ghosh22NH}
\begin{widetext}
\vskip -0.3cm
\begin{align}
H_0(\vect{k})&= \left( \lambda^s_1 \sin k_x +\lambda^s_2 \sin 2 k_x  \right) \Gamma_1 + \! \left( \lambda^s_1 \sin k_y \!+\!\lambda^s_2 \sin 2 k_y  \right)\Gamma_2 \! + \! \left[m_0-\lambda^h_1 \left(\cos k_x +\cos k_y\right) \! - \! \lambda^h_2 \left(\cos 2 k_x +\cos 2 k_y\right) \right] \! \Gamma_3 \non \\
&+\left[\lambda^f_1 (\cos k_x - \cos k_y)+\lambda^f_2 (\cos 2 k_x - \cos 2 k_y) \right] \Gamma_4 \ ,
\end{align}
\vskip -0.3cm
\end{widetext}
where, $\Gamma_1=\sigma_x s_z$, $\Gamma_2=\sigma_y s_0$, $\Gamma_3=\sigma_z s_0$, $\Gamma_4=\sigma_x s_x$. We consider the strengths of first (second) NN hopping, spin-orbit coupling, 
and $C_4$ symmetry breaking mass terms as $\lambda^h_1$, $\lambda^s_1$, and $\lambda^f_1$, ($\lambda^h_2$, $\lambda^s_2$, and $\lambda^f_2$), respectively. Here, $m_{0}$ is the staggered mass term. In what follows, we refer to the case of $\lambda_1\ne0$ and $\lambda_2 =0$ ($\lambda_{1,2}\ne0$) as first (second) NN model. 
In the absence of the second NN terms \ie~$\lambda^h_2,\lambda^s_2,\lambda^f_2 = 0$ and $\lambda^h_1,\lambda^s_1,\lambda^f_1 \neq  0$, the Hamiltonian $H_0(\vect{k})$ exhibits bulk gap closing at $m_0=\pm 2\lambda^h_1$. One can show that the Hamiltonian $H_0(\vect{k})$ exhibits four zero-energy corner modes arising within the regime $-2\lambda^h_1<m_0<2 \lambda^h_1$, manifesting a SOTI. Note that, in the absence of Wilson-Dirac mass terms $\lambda^f_{1,2} = 0$, one obtains gapless edge modes for $-2\lambda^h_1<m_0<2 \lambda^h_1$ except for $m_0=0$. This indicates the fact that the first-order topological insulator phase completely transforms into the second-order topological insulator phase. Furthermore, the presence of second NN terms, \ie~$\lambda^h_2,\lambda^s_2,\lambda^f_2 \neq 0$ and $\lambda^h_1,\lambda^s_1,\lambda^f_1 \neq  0$ substantially modifies the phase boundaries. To be specific, the bulk gap closes at $m_0=-2\lambda^h_2,0,2 \lambda^h_1+ 2 \lambda^h_2$. In this case, the system harbors four zero-energy corner modes for $-2\lambda^h_2<m_0<0$ while sixteen zero-energy corner modes for $0<m_0<2 \lambda^h_1+ 2 \lambda^h_2$. Similar to the earlier case, in the absence of Wilson-Dirac mass terms $\lambda^f_{1,2} = 0$, one obtains the gapless edge modes for $-2\lambda^h_2<m_0<2 \lambda^h_1+ 2 \lambda^h_2$ except for $m_0=0$.
%The mass term $m_0$ can be tuned to obtain bulk gap closings, which in turn induces topological phase transitions. In the absence of the second NN terms \ie $\lambda^f_1,\lambda^f_2 = 0$, the Hamiltonian $H_0(\vect{k})$ exhibits bulk gap closing at $m_0=\pm 2\lambda^h_1$. And one can show that the Hamiltonian $H_0(\vect{k})$ exhibits four zero-energy corner modes arising for $-2\lambda^h_1<m_0<2 \lambda^h_1$, manifesting a SOTI. Furthermore, the presence of second NN terms, \ie $\lambda^f_1,\lambda^f_2 \neq 0$ substantially modifies the phase boundaries. To be specific, the bulk gap closes when $m_0=-2\lambda^h_2,0,2 \lambda^h_1+ 2 \lambda^h_2$. In this case, the system harbors four and sixteen zero-energy corner modes for $-2\lambda^h_2<m_0<0$ and $0<m_0<2 \lambda^h_1+ 2 \lambda^h_2$, respectively.
Note that, this model also hosts first order topological insulator phase for $-2 \lambda^h_1<m_0<2 \lambda^h_1$ and $-2 \lambda^h_2<m_0<2 \lambda^h_1+ 2 \lambda^h_2$, respectively, when $(\lambda^f_1,\lambda^f_2)=(\ne 0,0)$ and $(\ne 0,\ne 0)$. The above model preserves  CS $C=\sigma_x s_y$ such that $C H_0(\vect{k}) C^{-1}=-H_0(\vect{k})$. Importantly, the time-reversal symmetry $T=i \sigma_0 s_y \mathcal {K}$ is broken when $\lambda^f_{1,2}\ne 0$ such that $T H_0(\vect{k}) T^{-1} \ne H_0(-\vect{k})$ with $\mathcal {K}$ being the complex-conjugation. Note that $T^2=-1$ leads to the AII class with a $\mathbb{Z}_2$ classification of the first-order topological phase.

%particle-hole symmetry (PHS) $P=A\mathcal {K} $ with $A=\sigma_x s_z$

%%%%%%%%%%%%%%%%%%%%% non Hermiticity and symmetries %%%%%%%%%%%%%%%%%%%%% 
Having demonstrated the physics of the Hermitian SOTI model, we now focus on the NH version of the above. We associate the  NH effect to the spin-orbit coupling part of the above Hamiltonian: $H_{\gamma}(\vect{k})=H_0(\vect{k})+ i \left( \gamma_x \Gamma_1+ \gamma_y \Gamma_2 \right)$ resulting in $H^{\dagger}_{\gamma}(\vect{k}) \ne H_{\gamma}(\vect{k})$. We consider $\gamma_x=\gamma_y=\gamma$ unless mentioned otherwise. The non-Hermiticity considered here can be thought of as an imaginary fictitious Zeeman field~\cite{YaoPRLSecond2018,Ghosh22NH}. %Notice that $H_{\gamma}(\vect{k})$ preserves PHS, generated by $A$ as follows $A H^*_{\gamma}(\vect{k}) A^{-1}=-H_{\gamma}(-\vect{k})$. 
The CS continues to be preserved as $C H_{\gamma}(\vect{k}) C^{-1}=-H_{\gamma}(\vect{k})$. We exploit the CS to define the NH analog of quadrupole moment in real space (see latter text for discussion). Note that, for a NH Hamiltonian, the CS is often also referred to as sublattice symmetry~\cite{KawabataPRX2019}. In the rest of the paper, we consider $\lambda^h_{1,2}=1$ for the sake of simplicity.   

%---------------------------------------------------------------------
%--------------------------------------------------------------------
\begin{figure}
	\centering
	\includegraphics[width=0.99\linewidth]{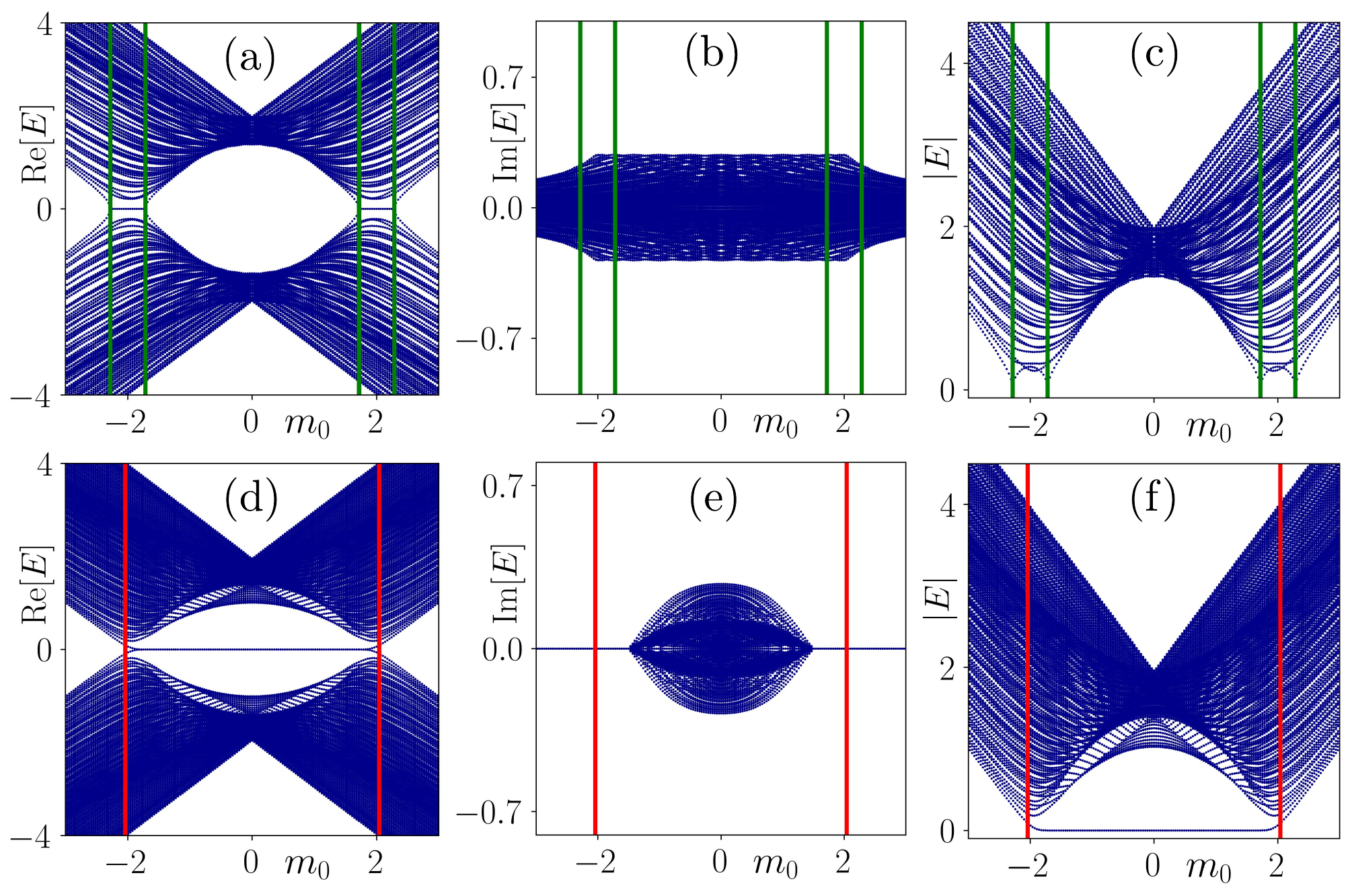}
	\caption{The energy spectra under PBC and OBC are illustrated as a function of the staggered mass term $m_{0}$ in upper and lower panels, respectively, for first NN hopping. The above panels 
	correspond to (a,d) Re[$E$], (b,e) Im[$E$] and (c,f) $|E|$. The model parameters are chosen as $\lambda^s_1=\lambda^h_1=\lambda^f_1=1.0$, and $\gamma=0.2$. Here, green lines correspond 
	to the exceptional points obtained from the PBC case $m_0=\pm 2 \pm \tilde{\gamma}$. The red lines, representing the exceptional phase boundary under OBC, are given by 
	$m_0=\pm (2 + \gamma^2_1)$.}
	\label{fig:NHEigenNN}
\end{figure}
%---------------------------------------------------------------------
%---------------------------------------------------------------------

%======================================================
\section{Results}\label{result}
%======================================================
In this section, we discuss the main results of this manuscript. We analyze the eigenvalue spectra and local density of states~(LDOS) corresponding to our NH model with first and second NN hopping. Afterward, we define the quadrupolar winding number for our system and demonstrate the phase diagram.

%%%%%%%%%%%%%%%%%%%%%%%%  complex energy in OBC for first NN model %%%%%%%%%%%%%%%%%%%%%%%
%----------------------------------------------------------
\subsection{NH Model with first NN hopping}\label{firstNN}
%----------------------------------------------------------

%%%%%%%%%%%%%%%%  energy with OBC %%%%%%%%%%%%%%%%%%%%%%
%%%%%%%%%%%%%%%%%%%%%%%%%%%%%%%%%%%%%%%%%%%%%%%%%%%%%%%%%

We begin with the energy dispersion of the NH SOTI model in the presence of the first NN, as shown in Figs.~\ref{fig:NHEigenNN} (a,b,c) [(d,e,f)] under periodic boundary condition (PBC) [(OBC)]. 
While employing PBC, we find ${\rm Re}[E]=0$ for   $ 2 -\tilde{\gamma} <m_0 <2 +\tilde{\gamma} $ and $ -2 -\tilde{\gamma} <m_0 <-2 +\tilde{\gamma} $ with $\tilde{\gamma}=\sqrt{2 \gamma^2}$ 
around $m_0=\mp 2$ as depicted by the gapless regions in Fig.~\ref{fig:NHEigenNN}(a), bounded by the green lines. These exceptional boundaries around $m_0=2 $, $-2 $, respectively, can be 
understood from the two-fold degeneracies of energy bands at $\vect{k}=(0,0)$ and $(\pi,\pi)$ such that $\lvert E(\vect{k}_{\rm EP})\rvert=0$. Interestingly, these bulk gapless exceptional points 
$m_0=\pm 2 \pm \tilde{\gamma} $ are exclusively noticed in $|E|$ under the PBC case as depicted in Fig.~\ref{fig:NHEigenNN}(c) by the green lines. 

We find that the complex energy spectra obtained under OBC and depicted in Figs.~\ref{fig:NHEigenNN}(d,e,f) do not mimic the underlying PBC nature. One can observe SOTI modes for which the 
real part of energy vanishes according to $-2 - \gamma_1^2<m_0<2 +  \gamma_1^2$ with $\gamma_1=  \gamma/\lambda^s_1$ as depicted by the red lines in Fig.~\ref{fig:NHEigenNN}(d). 
These boundaries can be anticipated by the non-Bloch form of momentum $k_i\to k'_i - i \gamma/\lambda^s_1$ with $i=x,y$ where the renormalized mass term $m'_0=m_0-2- \gamma_1^2<0~(>0)$  determines the topological (trivial) phase of the NH model~\cite{YaoPRLSecond2018,Ghosh22NH}. Note that, for Bloch momentum $\vect{k}=(0,0)$ [$\vect{k}=(\pi,\pi)$], the exceptional phase boundaries extend till $m_0=  \pm (2 +  \gamma_1^2)$ leading to the emergence of exceptional SOTI phases beyond the Hermitian gapless phase boundaries $m_0=  \pm 2$.   
%Interestingly, imaginary parts of the energy associated with all the single particle states vanish for $m_0> |2 \lambda^h_1|$. 
All the single-particle energy states under OBC except the corner modes exhibit an imaginary component of energy for $|m_0|<2 $ as shown in Fig.~\ref{fig:NHEigenNN}(e). 
This is markedly different from the PBC case, depicted in Fig.~\ref{fig:NHEigenNN}(b), where single particle states have finite amount of imaginary energy for $|m_0|>2$.
This refers to a macroscopic degeneracy within a certain range of $m_0$ as far as the ${\rm Im}[E]$ is considered. Since such macroscopic degeneracy does not exist for ${\rm Re}[E]$, the $|E|$ demonstrates the NH corner modes for $|m_0|<|2  +  \gamma_1^2|$ under OBC [see red lines in Fig.~\ref{fig:NHEigenNN}(f)], while non-Hermiticity mediated bulk gapless points are noticed for the 
PBC case, see  Fig.~\ref{fig:NHEigenNN}(c).

%---------------------------------------
\begin{figure}
	\centering
	\includegraphics[width=1.04\linewidth]{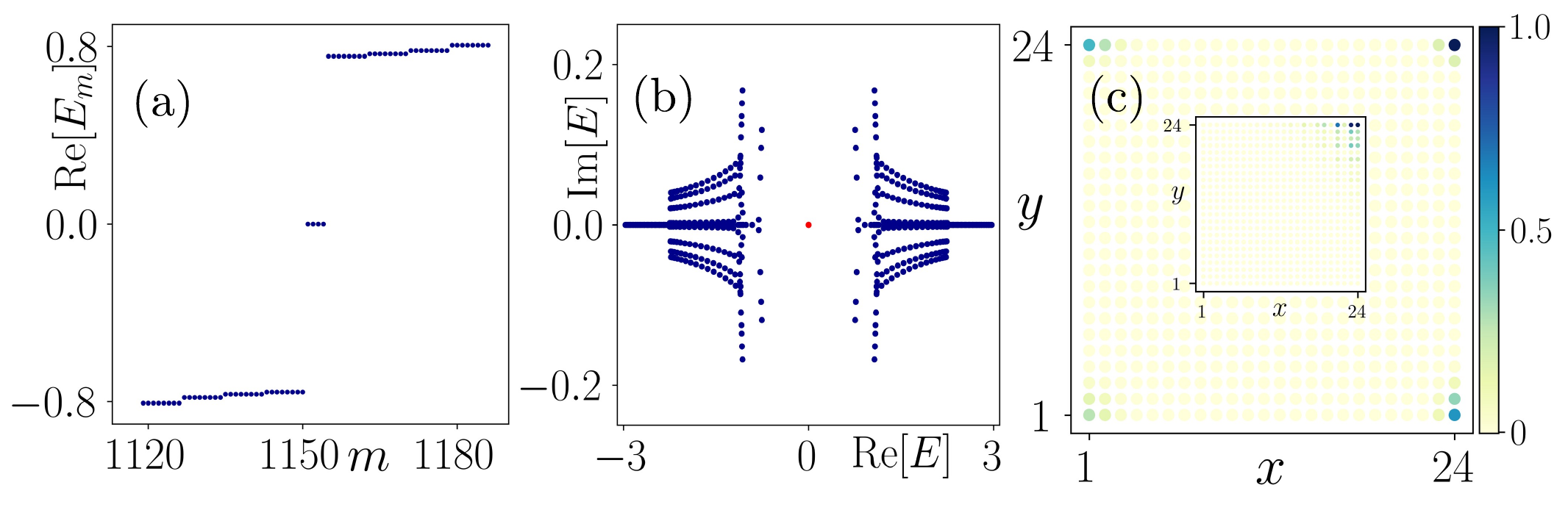}
	\caption{(a) The real part of the energy eigenvalue spectrum ${\rm Re}[E_m]$ obtained under OBC is shown as a function of the state index $m$. (b) The eigenvalue spectrum in the ${\rm Re}[E]
	\mhyphen{\rm Im}[E]$ plane is illustrated. The eigenvalues corresponding to the corner state are marked by the red dots. (c) The LDOS spectrum associated with the ${\rm Re}[E]=0$ is depicted in the 2D domain. In the inset, we demonstrate the LDOS associated with a bulk state with $E=-2.052543$.} We choose $m_0=1.0$, while the other model parameters remain the same as mentioned in Fig.~\ref{fig:NHEigenNN}.
	\label{fig:NHELDOSNN}
\end{figure}
%---------------------------------------

Having understood the generation of the NH SOTI phase as a function of the topological mass $m_0$, we consider a slice with $m_0=1$ from Fig.~\ref{fig:NHEigenNN} and analyze the results presented 
in Fig.~\ref{fig:NHELDOSNN}. In particular, employing OBC, we depict the real part of the eigenvalue spectrum ${\rm Re}[E_m]$ close to ${\rm Re}[E]=0$ as a function of the state index $m$ in 
Fig.~\ref{fig:NHELDOSNN}(a).%, for a system obeying OBC. 
We observe the appearance of four states at ${\rm Re}[E]=0$, which corresponds to localized corner states. In Fig.~\ref{fig:NHELDOSNN}(b), we illustrate the eigenvalue spectrum in the ${\rm Re}[E]\mhyphen{\rm Im}[E]$ plane. The corner modes are marked by the red dot, which indicates that the corner modes have both real and imaginary parts of the eigenvalue equal to zero.
The CS of the model is reflected in the symmetric profile of energy on the positive and negative sides of the real energy. The corner states (red) are clearly separated from the other states (blue) by a line gap at ${\rm Re}[E]=0$. Moreover, we show the site-resolved 
LDOS distribution in Fig.~\ref{fig:NHELDOSNN}(c). We find that the corner modes are mostly localized at only one corner of the 2D domain. This phenomenon of the localization of the corner modes 
limited to only one corner of the system has been investigated previously in NH higher-order systems where mirror symmetries play a crucial role~\cite{NoriNHPRL2019,Ghosh22NH}. In particular, the localization of the states at a single corner of the 2D domain can attributed to the mirror rotation symmetry $M_{xy}$. By changing the signs of the NH terms $\gamma_x$ and $\gamma_y$, we can change the location of the corner states. By breaking the mirror rotation symmetry for $\gamma_x\neq\gamma_y$, we can localize the corner states at more than one corner. We discuss different spatial symmetries of the system and shifting of the corner modes in Appendix~\ref{App:A}. In the inset of Fig.~\ref{fig:NHELDOSNN}(c), we also show the LDOS distribution associated with a bulk state. We observe that the bulk state is also localized at the corner of the system, indicating the existence of higher-order skin-effect~\cite{Kawabata20b}.

%--------------------------------------------------------------
\subsection{NH Model with second NN hopping}\label{secondNN}
%--------------------------------------------------------------

%%%%%%%%%%%%%%%%%%%%% energy with  NNN %%%%%%%%%%%%%%%%%%
%%%%%%%%%%%%%%%%%%%%%%%%%%%%%%%%%%%%%%%%%%%%%%%%%%%%%%%%%
%----------------------------------------------------------------------------------------------------------------
%----------------------------------------------------------------------------------------------------------------
\begin{figure}[]
	\centering
    \includegraphics[width=0.99\linewidth]{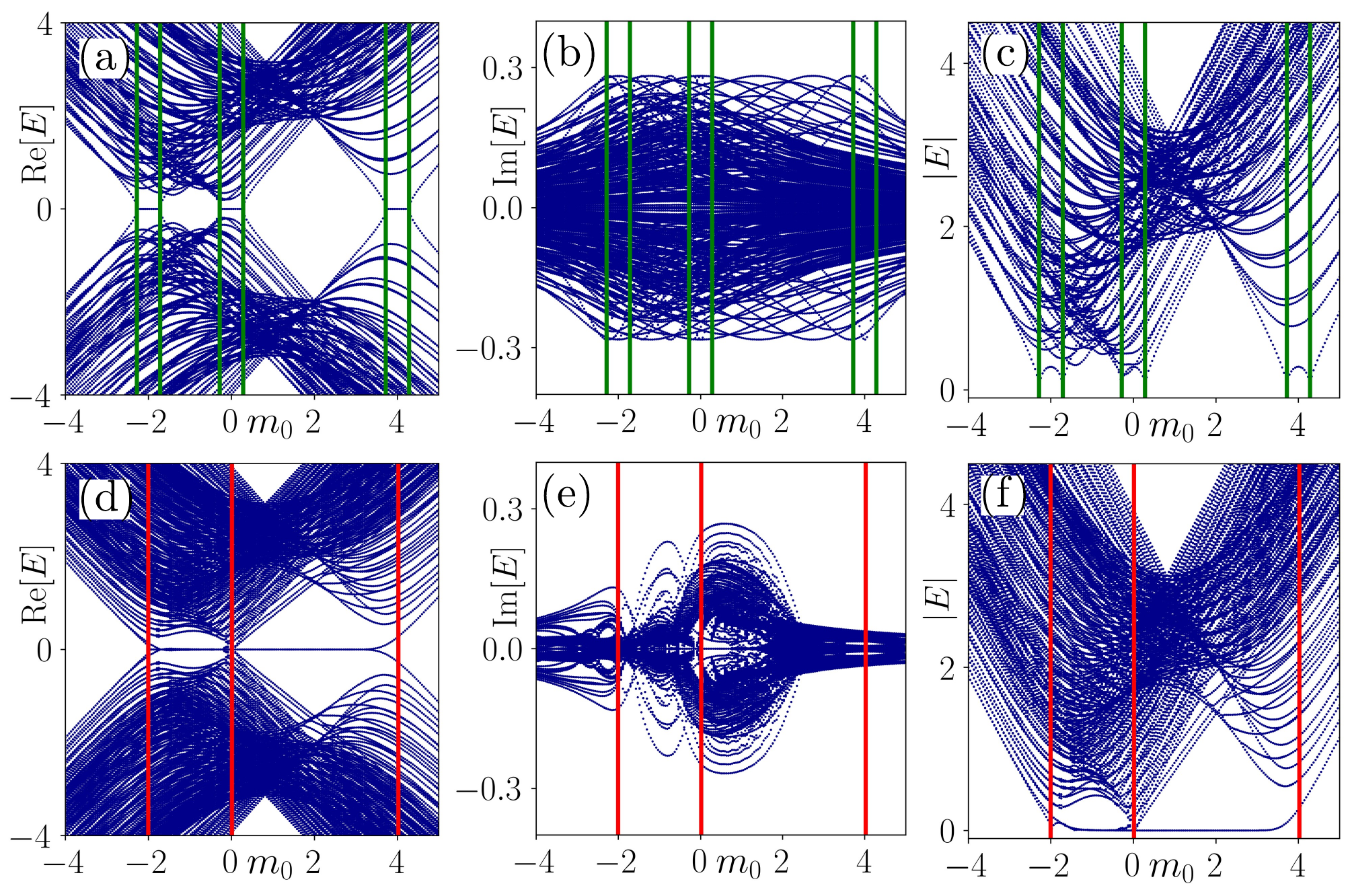}	
    \caption{The energy spectra under PBC and OBC are illustrated in upper and lower panels, respectively as a function of $m_{0}$, for second NN hopping. The panels correspond to (a,d) Re[$E$] 
    (b,e) Im[$E$] and (c,f) $|E|$. We choose the model parameters as $\lambda^s_1=\lambda^h_1=\lambda^f_1=1.0$, $\lambda^s_2=\lambda^h_2=\lambda^f_2=1.0$ and $\gamma=0.2$. The green lines 
    correspond to the exceptional points $m_0= 4  \pm\tilde{\gamma}$, $\pm\tilde{\gamma}$, and $-2 \pm\tilde{\gamma}$ under PBC. The red lines, representing the exceptional phase boundary 
    under OBC, are given by $m_0= 4 + 5  \gamma_2^2$, $3\gamma_2^2 $, and $-2 -\gamma_2^2$.}
	\label{fig:NHEigenNNN}
\end{figure}
%---------------------------------------------------------------------------------------------------------------
%---------------------------------------------------------------------------------------------------------------
%%%%%%%%%%%%%%%%%%%%%%%%%%%%%%%%%%%%%%%%%%%%%%%%%%%%%%%%%

To start with, we depict the energy dispersion of the NH SOTI model in the presence of the second NN in Figs.~\ref{fig:NHEigenNNN} (a,b,c) [(d,e,f)] under PBC [OBC]. We find degenerate eigenstate 
with ${\rm Re}[E]=0$ under PBC for $ -2 - \tilde{\gamma}< m_0< -2 + \tilde{\gamma}$, $  - \tilde{\gamma}< m_0<  \tilde{\gamma}$, and $ 4 - \tilde{\gamma}< m_0< 4 + \tilde{\gamma}$ as depicted 
by the green lines in Fig.~\ref{fig:NHEigenNNN}(a). %as depicted by green lines. 
These exceptional boundaries around $m_0=-2 $, $0$ and $4$, respectively, can be understood from the two-fold degeneracies of energy bands at $\vect{k}=(\pm 2\pi/3,\pm 2\pi/3)$, $(\pi,\pi)$, 
and $(0,0)$ such that $\lvert E(\vect{k}_{\rm EP})\rvert=0$. Similar to the earlier first NN model, these bulk gapless exceptional points $m_0=-2  \pm \tilde{\gamma} $, $ \pm \tilde{\gamma} $, and 
$4 \pm \tilde{\gamma} $ are exclusively observed in $|E|$ for PBC case as depicted in Fig.~\ref{fig:NHEigenNNN}(c) by green lines. 

We now examine the energy spectrum for the second NN model in Figs.~\ref{fig:NHEigenNNN}(d,e,f) employing OBC. Similar to the first NN case, the momentum takes the following non-Bloch form 
$k_i\to k'_i - i \gamma_2$ with $\gamma_2=  \gamma/(\lambda^s_1+2\lambda^s_2) $ and $i=x,y$. This leads to the renormalized mass term $m'_0=m_0-4 - 5 \gamma_2^2 <0~(>0)$ for the topological (trivial) phase with zero-energy (finite energy bulk) modes considering the Bloch momentum $\vect{k}=(0,0)$. On the other hand, another topological (trivial) phase with zero-energy (finite energy) modes appears for $-m'_0=m_0+2 + \gamma_2^2>0~(<0)$ while exploiting energy around the Bloch momentum $\vect{k}=(\pm 2\pi/3,\pm 2\pi/3)$. Our analysis indicates the existence of four [sixteen] corner modes for $3 \gamma_2^2 <m_0< 4  + 5  \gamma_2^2$ [$-2 - \gamma^2_2<m_0<3 \gamma_2^2$] yielding ${\rm Re}[E]=0$ under OBC. However, the numerical findings shown in Fig.~\ref{fig:NHEigenNNN}(d) and the topological regime highlighted by the red lines do not fully match with the exceptional boundaries predicted analytically. This can be attributed to the finite size effect in the second NN case, which is substantially small for the first NN case. In particular, the rightmost boundary in Fig.~\ref{fig:NHEigenNNN}(d) is more affected due to the finite size scaling as there exists a significant mismatch between the analytical prediction $m_0= 4  + 5  \gamma_2^2$ and real zero-energy modes obtained under OBC. To this end, we discuss the finite size scaling around the phase transition point at the rightmost part of Fig.~\ref{fig:NHEigenNNN}(d) in Appendix~\ref{App:B}. Note that, the finite size effect is expected to become more substantial in case of OBC rather than for PBC. Due to this reason, we choose the phase boundary at $m_0= 4  + 5  \gamma_2^2$ for the finite size analysis. We find macroscopic degeneracies at ${\rm Re}[E]=0$ around $m_0= 0$, unlike the previous case. The complex energy spectrum ${\rm Im}[E]$ does not manifest any noteworthy features, as shown in Figs.~\ref{fig:NHEigenNNN}(b,e), irrespective of PBC and OBC cases. The absolute value of energy is expected to vanish \ie $|E|=0$ under OBC for $-2-  \gamma^2_2<m_0<4 + 5 \gamma_2^2 $. However, numerical results suffer from finite size effects as far as the exceptional boundaries are concerned [see the red lines in Fig.~\ref{fig:NHEigenNNN}(f)]. The finite value of $\gamma$ (NH effect) thus extends the Hermitian topological phase beyond its boundaries, leading to exceptional topological phases. 

%--------------------------------------------------------------------------------------------------------------
%--------------------------------------------------------------------------------------------------------------
\begin{figure}[]
	\centering
	\includegraphics[width=1.03\linewidth]{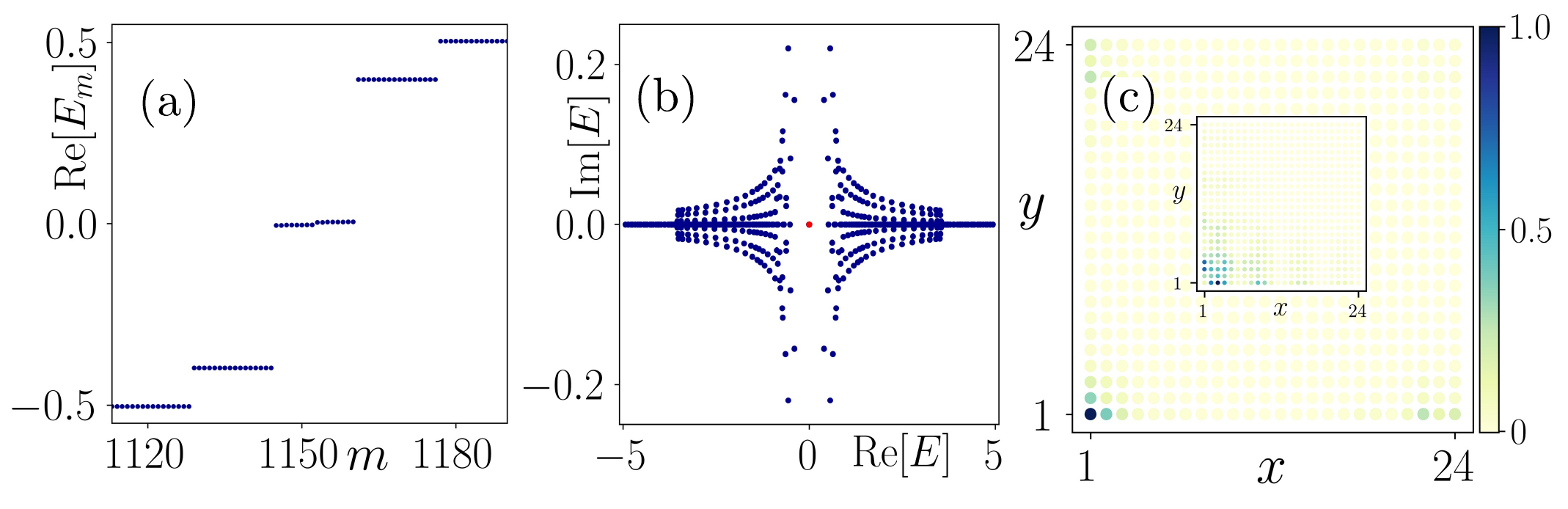}
	\caption{(a) The real part of the energy spectrum ${\rm Re}[E_m]$ obtained under OBC is shown as a function of the state index $m$. (b) The eigenvalue spectrum in the ${\rm Re}[E]\mhyphen{\rm Im}[E]$ plane is demonstrated. The corner state eigenvalues are indicated by the red dot. (c) The LDOS associated with the ${\rm Re}[E]=0$ is depicted in the 2D lattice. In the inset, we illustrate the LDOS associated with a bulk state with $E=-3.291226$. We choose $m_0=-1.0$, while the other model parameters remain the same as mentioned in Fig.~\ref{fig:NHEigenNNN}.}
	\label{fig:NHELDOSNNN}
\end{figure}
%-------------------------------------------------------------------------------------------------------------
%-------------------------------------------------------------------------------------------------------------

Moreover, we also analyze the eigenvalue spectrum, choosing a fixed value of $m_0$. As discussed before, we obtain four corner states when $3 \gamma_2^2<m_0<4 + 5 \gamma_2^2 $. The corresponding eigenvalue spectrum and LDOS, when the system exhibits four corner states, remain qualitatively the same as depicted in Fig.~\ref{fig:NHELDOSNN} for the first NN case. Thus, we do not repeat that analysis here. Rather, we choose the value of $m_0$ in such a way that we obtain sixteen corner states. For this case, we show the real part of the eigenvalue spectrum ${\rm Re}[E_m]$ close to ${\rm Re}[E]=0$ as a function of the state index $m$ in Fig.~\ref{fig:NHELDOSNNN}(a), for a system obeying OBC. One can note the existence of sixteen states at ${\rm Re}[E]=0$, which corresponds to localized corner modes. The finite separation from the exact ${\rm Re}[E]=0$ can be attributed to the finite size effect. Nevertheless, in Fig.~\ref{fig:NHELDOSNNN}(b), we illustrate the eigenvalue spectrum in the ${\rm Re}[E]\mhyphen{\rm Im}[E]$ plane where the CS manifests itself in the symmetric profile of real energy. The corner modes are marked by the red dot. The line gap feature behaves in a similar way to the previous case. %is observed here as well like the previous case. 
The corner modes have both real and imaginary parts of the eigenvalue equal to zero. We also illustrate the site-resolved LDOS distribution in Fig.~\ref{fig:NHELDOSNNN}(c). We find that the 
corner modes are mostly localized at only one corner of the 2D system, similar to the first NN hopping case. Furthermore, we also observe the signature of higher-order skin effect of the bulk states. To highlight this, we depict the LDOS associated with a bulk state in the inset of Fig.~\ref{fig:NHELDOSNNN}(c). It is evident that the bulk state is localized at the corners of the system.

%----------------------------------------------------
\subsection{Quadrupole Winding number}
\label{qwn}
%----------------------------------------------------
%%%%%%%%%%  Quadrupole Winding number %%%%%%%%%%%%%%%%%

We now investigate the topological invariant, namely, quadrupole winding number (QWN), by exploiting the CS. In the following discussion, we first illustrate the Hermitian version of QWN. Given the 
fact that CS constraints $C H_{0}(\vect{k}) C^{-1} = -H_{0}(\vect{k})$, we can anti-diagonalize the Hamiltonian in the basis of the CS operator spanned by $U_C$ as follows~ \cite{ryu2010topological,Shinsei16}
\begin{equation}
	\tilde{H}_0=U_C^\dagger H_0 U_C= \begin{pmatrix}
		0 & h \\
		\tilde{h} & 0
	\end{pmatrix} \ ,
\end{equation}
Here, $\tilde{h}=h^\dagger$ if $\tilde{H}_0$ is Hermitian. We find $U_C C U^{\dagger}_C=\pm 1$, suggesting that CS can be classified into two kinds of sub-lattices, namely $A$ and $B$ for $+$ and $-$ expectation values, respectively. This further entails that $U_C= U^A_C -U^B_C$ where $U^A_C=\sum_{\alpha \in A}|\alpha\rangle \langle \alpha|$ and $U^B_C=\sum_{\beta \in B}|\beta\rangle \langle \beta|$.   

Employing singular value decomposition of $h$, we obtain $h=U_A \Sigma U_B^\dagger$ where $U_{A,B}$ are unitary matrices and $\Sigma$ denotes a diagonal matrix. Note that, the diagonal elements of $\Sigma$ are referred to as singular values. One can compute the flattened Hamiltonian $Q$, having eigenvalue $\pm 1$ as follows~\cite{Benalcazar22}
\begin{equation}
	Q= \begin{pmatrix}
		0 & q \\
		q^{\dagger} & 0
	\end{pmatrix}\ ,
\end{equation}
with $q=U_A U^{\dagger}_B$ being a unitary matrix. It has been shown that the winding number, derived using $q$ and $q^{\dagger}$, is related to the relative polarization of $A$ and $B$ sublattice. In a similar spirit, the winding number in the real space is given by \cite{Lin21}
\begin{equation}
\label{eqn:winding_number_rs}
\nu  = \frac{1}{{2\pi i}}{\rm{Tr}}\left[ {\log \left( {{{ \mathcal{X}}_A} \mathcal{X}_B^{\dagger}} \right)} \right]\ ,
\end{equation}
where, ${{\mathcal{X}}_\sigma } = U_\sigma ^{\dagger} U^{\sigma}_C \mathcal{X} U^{\sigma}_C {U_\sigma }$ ($\sigma=A,B$) are unitary matrices. The operator ${{ \mathcal{X}}_\sigma }$ denotes the sublattice dipole operator, which is the projection of the position operator onto the $\sigma$ sector of the chiral basis. The position operator, \ie the dipole operator
$\mathcal{X}=\exp(2i \pi x/L)$ is defined on a periodic array of one-dimensional length $L$.  

Now turning to the two-dimensional system where the dipole operator $\mathcal{X}$ can be replaced by the quadrupole operator $\mathcal{Q}=\exp(2i\pi x y/L_x L_y)$. This results in the sublattice quadrupole operator $\mathcal {Q}_{\sigma}=U_\sigma^{\dagger} U^{\sigma}_C \mathcal{Q} U^{\sigma}_C U_{\sigma}$. Therefore, the QWN can be defined as~\cite{Benalcazar22}
\begin{equation}
	N_{xy}=  \frac{1}{2 \pi i} {\rm Tr} \left[\log\left( \mathcal{Q}_A\mathcal{Q}^{\dagger}_B\right)\right]\ ,
\end{equation}
This invariant is quantized to an integer number and predicts the number of topologically protected corner states at each corner of the 2D lattice.

%%%%%%%%%%%% NH extension of QWN %%%%%%%%%%%%%%%%%%%%%%%%%%%%%%%%
We now examine the present situation with $\gamma\ne 0$ where the NH analog of QWN is discussed. Importantly, CS is also preserved for the NH Hamiltonian  $C H_{\gamma}(\vect{k}) C^{-1} = -H_{\gamma}(\vect{k})$ allowing for the anti-diagonal form of  $\tilde{H}_{\gamma}$. At the same time, the definition of  $U^{A,B}_C$ remains unaltered for the NH case. We adopt the bi-orthogonalized definitions of $U_A$ and $U^{\dagger}_B$, obtained from the singular value decomposition of $h$, to define $U^{\dagger}_A$ and $U_B$, respectively.  One has to ensure $\sum_{n}|U^{R}_{\sigma,n}\rangle \langle U^{L}_{\sigma,n}|=\mathds{1}$ and $\langle U^{L}_{\sigma,n}| U^{R}_{\sigma,m}\rangle =\delta_{mn}$ with $\sigma=A,B$ and $L(R)$ denotes the left (right) singular vectors. This results in left [right] singular vectors corresponding to right singular vectors $(U^{\dagger}_B)^{\dagger}\equiv U^R_B$ [left singular vectors $U^{\dagger}_A \equiv U^L_A$]  as $ U^{\dagger}_B \equiv U^{L}_B$ [$U_A \equiv U^{R}_A$]. Therefore, sublattice quadrupole operator $\mathcal {Q}_{\sigma}$ takes the form $\mathcal {Q}_{\sigma}=U_\sigma^{L} U^{\sigma}_C \mathcal{Q} U^{\sigma}_C U^{R}_{\sigma} $. In addition, the non-Bloch form of momentum has to be incorporated while computing $\mathcal {Q}_{\sigma}$. 

To be precise, the complex momentum $k_i\to k'_i - i \gamma_2$ with $i=x,y$ leads to the exponentially enhanced and suppressed hopping elements by the multiplicative factors $\exp(\gamma_1)$ and 
$\exp(-\gamma_1)$ [$\exp(\gamma_2)$ and $\exp(-\gamma_2)$] for first [second] NN models. We use the real space form of the tight-binding model with the renormalized hopping amplitudes as follows 
$\lambda^{s,h,f}_{1,2} \to \lambda^{s,h,f}_{1,2} \exp(\gamma_{1,2})$ and $\lambda^{s,h,f}_{1,2} \to \lambda^{s,h,f}_{1,2} \exp(-\gamma_{1,2})$ for forward and backward hopping amplitudes, respectively. 
We consider the real space version of NH Hamiltonian $H_{\gamma}$ with the above-mentioned renormalized hoppings in order to compute QWN. Altogether, this enables us to define the NH analog of 
QWN $N_{xy}$ with dressed hopping and bi-orthogonalized definition. Note that the real part of $N_{xy}$ exhibits quantized value for the present case with $\gamma \ne 0$ as demonstrated below.

%%%%%%%%%%%%%%%%%%%%%%% phase diagram in NH-mass plane for first and second NN model %%%%%%%%%%%%%

\begin{figure}[]
    \centering
    \includegraphics[width=1.03\linewidth]{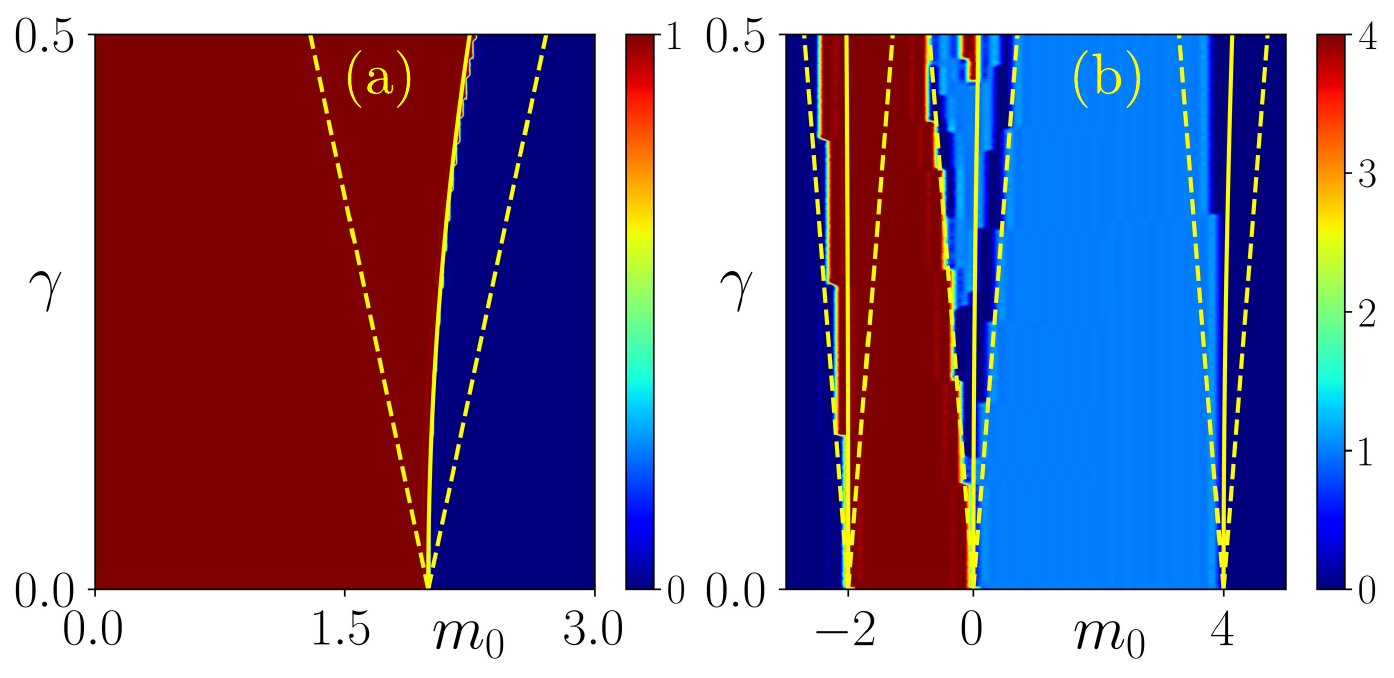}    
    \caption{The phase diagram in $m_0$-$\gamma$ plane for NH first NN and second NN hopping models are shown in panel (a), and (b), respectively. The color bar represents QWN $N_{xy}$. 
    The phase boundary $m_0=2 + \gamma_1^2$  [$m_0=2 \pm \tilde{\gamma}$], obtained from OBC [PBC], between exceptional topological and trivial phases in panel (a) is identified by the yellow 
    solid [dashed] line. The phase boundaries, associated with OBC [PBC], between topological phases hosting sixteen and four corner modes and trivial phase, are given by $m_0= 4  + 5 \gamma_2^2$ 
    [$m_0=4 \pm \tilde{\gamma}$], $m_0= 3 \gamma_2^2$ [$m_0=\pm \tilde{\gamma}$], and $m_0= -2  - \gamma_2^2$ [$m_0=-2 \pm \tilde{\gamma}$], respectively, represented by the yellow 
    solid [dashed] lines in panel (b).}
    \label{fig:nu}
\end{figure}

We now discuss the phase diagram of the first and second NN NH model in the $\gamma$-$m_0$ plane in Figs.~\ref{fig:nu}(a,b) respectively. Note that, there exists four corner modes with 
${\rm Re}[E]=0$ for $m_0< 2 + \gamma_1^2$, yielding $N_{xy}=1$. This refers to the fact that there exists only one topological zero-mode per corner [see Fig.~\ref{fig:nu}(a)]. On the other hand, when $m_0> 2 +  \gamma_1^2$, the NH model does not host any topological phase and hence $N_{xy}=0$. Hence, the topological phase boundary is $m_0= 2 +  \gamma_1^2$ which is indicated by the yellow solid line. This is also predicted from the complex energy spectrum with OBC. The yellow dashed lines in Fig.~\ref{fig:nu}(a) represent $m_0 =2 \pm \tilde{\gamma}$ lines as predicted from the complex energy under PBC. Interestingly, the real space invariant QWN fails to identify these phase boundaries. Unlike the Hermitian system, the phase boundaries between topological and trivial phases cannot 
be captured by the energy spectrum under OBC and PBC in the present NH case. This clearly suggests that the topological phase, predicted from the complex energy spectrum in OBC, is apprehended 
by the non-Bloch and bi-orthogonalized version of QWN. We note that the analytically derived phase boundaries are valid for $\gamma_{1,2} \ll 1$.  Interestingly, the non-Hermiticity induces additional regions $-2 - \gamma_1^2 <m_0<-2$ and  $2 <m_0<2+ \gamma_1^2$ around $m_0=\pm 2$ beyond the Hermitian phase boundaries. In Fig.~\ref{fig:nu}(a), we only illustrate the positive $m_0$ window where the NH SOTI phase is present for $m_0<2+ \gamma_1^2$.

We find qualitatively similar results in the case of the second NN model, as shown in Fig.~\ref{fig:nu}(b).  In addition to $N_{xy}=1$ phase, we obtain $N_{xy}=4$ phase where four zero-energy modes 
with ${\rm Re}[E]=0$ are present at each corner. While investigating the phase boundaries, it is expected to find $N_{xy}=1$ [$4$] for $3 \gamma_2^2  <m_0< 4  + 5 \gamma_2^2$ 
[$-2 - \gamma^2_2<m_0<3 \gamma_2^2$]. For smaller strength of non-Hermiticity \ie $\gamma \to 0$, we find quantitative agreement between the analytical and numerical findings. 
The phase boundaries $m_0=4  + 5 \gamma_2^2$, $3 \gamma_2^2 $,  and $-2 - \gamma^2_2$, designated by the yellow solid line, do not fully comply with the $N_{xy}$ profile for $\gamma>0.1$. 
This can be due to more intricacies than just the finite-size effect. Interestingly, the following tendency is noticed: for $\tilde{\gamma}  <m_0< 4  + 5 \gamma_2^2$, one obtains $N_{xy}=1$ while within
the regime $ -2 -\tilde{\gamma} <m_0< - \tilde{ \gamma}$, $N_{xy}$ acquires the value $4$. Therefore, the non-Bloch and bi-orthogonalized version of QWN can quantitatively 
and qualitatively identify the phase boundaries of SOTI phases hosting four and sixteen corner modes starting from the trivial phases for $m_0 >  4  + 5 \gamma_2^2$ and $m_0<-2 - \tilde{\gamma}$, respectively, across which $N_{xy}$ jumps between zero and finite values.  The real space invariant QWN is thus a useful topological marker to identify the exceptional phases for our NH system with OBC.

%%%%%%%%%%%%%%%%%%%%%%%%%%%%%%%%%%%%%%%%%%%%%%%%%%%%%%%%%%%%%%%%%%%%%%%%%%%%%%%%%%%%%%%%%%%%%%%%%%%%%%%%%%%%%%%%
%---------------------------------------------------------------------------------------------
%---------------------------------------------------------------------------------------------
%%%%%%%%%%%%%%%%%%%%%%%%%%%%%%%%%%%%%%%%%%%%%%%%%%%%%%%%%%%%%%%%%%%%%%%%%%%%%%%%%%%%%%%%%%%%%%%%%%%%%%%%%%%%%%%%
%--------------------------------------------------------------------------------------------
%--------------------------------------------------------------------------------------------

As mentioned earlier for the first NN case, the Hermitian phase boundaries are modified due to the non-Hermiticity. For second NN Hermitian counterpart with $\gamma=0$, one obtains SOTI phases $0 <m_0< 4 $ and $ -2  <m_0< 0$ hosting four and sixteen zero-energy corner modes, respectively. The NH factor $\gamma$ introduces four corner modes for positive values of $m_0$ beyond $m_0=4$ until $m_0< 4  + 5 \gamma_2^2$, as demonstrated in Fig.~\ref{fig:nu}(b). The same applies to the negative values of $m_0$ where the sixteen NH corner modes continue to exist for $|m_0|<2+\tilde{\gamma}$ beyond $m_0=-2$. Importantly, in the second NN SOTI model, the number of corner modes changes for positive and negative values of $m_0$ which is not the case for the first NN SOTI model. Therefore, we would like to emphasize that the NH factor $\gamma \neq 0$ and the second NN hoppings $\lambda^h_2,\lambda^s_2,\lambda^f_2 \neq 0$, together modify the phase diagram in a complex manner such that %of the model where 
the number of corner modes and their corresponding parameter window vary significantly as compared to the first NN Hermitian model. It would be interesting to study in future why the changes in QWN do not always follow the OBC energy gap closing lines. For example, $m_0=-2 -\gamma_2^2$ and $m_0=-2 -\tilde{\gamma}$ ($m_0= 3 \gamma_2^2$ and $m_0=\pm \tilde{\gamma}$) phase boundaries around $m_0=-2$ ($m_0=0$) can be investigated further for better understanding of the interplay between 
NH term and second NN hoppings.

On the other hand, between two SOTI phases with different number of corner modes (for $ -\tilde{\gamma} <m_0< \tilde{\gamma}$), we find $N_{xy}\ne 0$ as depicted in Fig.~\ref{fig:nu}(b). 
In the complex energy analysis under OBC, we find macroscopic degeneracies with ${\rm Re}[E]=0$ for $-2  - \gamma^2_2 <m_0<\tilde{\gamma}$. Likewise, the earlier first NN case, $N_{xy}$ does not exhibit any jumps between finite and zero values around $m_0=4\pm \tilde{\gamma}$ and $m_0=-2\pm \tilde{\gamma}$ which are predicted by the complex energy spectrum under PBC.
On the contrary, $m_0= \pm\tilde{\gamma} $ boundaries, predicted by complex energy spectrum under PBC, are visible as $N_{xy}$ changes between two finite values. The exceptional lines $m_0=-2\pm \tilde{\gamma}$, $\pm \tilde{\gamma}$, and $4 \pm \tilde{\gamma}$ obtained employing PBC are depicted by yellow dashed lines in Fig.~\ref{fig:nu}(b). This agreement is surprising and yet to be explored in the future. However, there is an apparent discrepancy between the solid yellow lines and the numerically obtained $N_{xy}$. This mismatch is owing to the fact that the mathematical form of non-Bloch transformation that we consider in the hopping terms while computing $N_{xy}$, employing PBC, is computed by employing a low-energy version of $H_{\gamma}(\vect{k})$. To obtain the low-energy spectrum of $H_{\gamma}(\vect{k})$, we expand the Hamiltonian around $\vect{k}=(0,0)$. By doing that, we can obtain the phase boundary associated with the right part of Fig.~\ref{fig:nu}(b). 
However, when we incorporate the second NN hopping elements, the low-energy model around $\vect{k}=(0,0)$ does not necessarily encapsulate all the phase transition lines. In that scenario, 
one should also consider a low-energy Hamiltonian around other momenta such as  $\vect{k}=(\pm 2 \pi/3, \pm 2 \pi/3)$, depending upon the value of $m_0$. Nevertheless, this scenario adds substantial complexity to the problem as one should consider a different non-Bloch form for different $m_0$. Thus, finding a universal transformation to obtain the exact phase boundary in the NH second NN hopping case still remains an interesting question and is beyond the scope of the present paper. Nevertheless, complex energy spectra under PBC might be useful for understanding the phase boundaries between 
two different topological phases.

%------------------------------------------------------------
%------------------------------------------------------------
\begin{table*}[ht]
%\textcolor{red}{
\begin{tabular}{|l|l|l|}
\hline
%\rowcolor[HTML]{C0C0C0} 
{ Symmetry} & { Operation}  & { Remarks}  
\\ \hline   
Mirror-$x$ & $M_x=  \sigma_0 s_y$: $M_x H_\gamma(k_x,k_y) M_x^{-1}= H_\gamma(-k_x,k_y)$ & Broken, when $\lambda^f_{1,2},\gamma_x \neq0$  
\\ \hline 
Mirror-$y$ & $M_y=  \sigma_z s_x$: $M_y H_\gamma(k_x,k_y) M_y^{-1}= H_\gamma(k_x,-k_y)$ & Broken, when $\lambda^f_{1,2},\gamma_y \neq0$ 
\\ \hline 
Four-fold rotation  & $C_4=e^{-\frac{i \pi}{4}\sigma_z s_z}$: $C_4 H_\gamma(k_x,k_y)C_4^{-1}= H_\gamma(-k_y,k_x) $  & Broken, when $\lambda^f_{1,2},\gamma_{x,y}\neq0$ 
\\ \hline 
Mirror-rotation I & $M_{xy}=C_4 M_y$: $M_{xy} H_\gamma(k_x,k_y) M_{xy}^{-1}= H_\gamma(k_y,k_x)$  &  Broken, when $\gamma_{x}\neq \gamma_{y}$ 
\\ \hline
Mirror-rotation II & $M_{x\bar{y}}=C_4 M_x$: $M_{x\bar{y}} H_\gamma(k_x,k_y) M_{x\bar{y}}^{-1}= H_\gamma(-k_y,-k_x)$  &  Broken, when $\gamma_{x}\neq \gamma_{y}$
 \\ \hline
\end{tabular}
\caption{Spatial symmetries and their operations are highlighted.}
\label{Table:symmetries}
%}
\end{table*}
%--------------------------------------------------------------
%--------------------------------------------------------------

%---------------------------------------------------------
\section{Summary and Conclusion}\label{conclusion}
%---------------------------------------------------------

To summarize, in this article, we consider a second NN hopping model in the presence of non-Hermiticity to investigate the emergence of second-order topological phases. By exploring the real part of the complex energy spectrum for the first and second NN NH models under OBC, we find that the former model only hosts four zero-energy corner modes while the latter model can host four as well as sixteen zero-energy corner modes as the hallmark of the NH SOTI phases. We compute the real space invariant, namely, bi-orthogonalized QWN, by keeping in mind the non-Bloch form of the momentum to uniquely characterize the different topological phases. The phase boundaries captured by the above invariant can successfully mimic the emergence of NH SOTI phases out of the trivial phases, as demonstrated by the complex energy dispersion under OBC for both the first and second NN models. The topological phase boundary between two different topological phases, observed in the second NN model, can be anticipated from the complex energy spectrum under PBC for the above model. In the future, one can include disorder to study the exceptional topological Anderson insulators hosting corner modes where a generalized version of the presently adopted real space topological index will have to be examined.

%======================================================
\subsection*{Acknowledgments}
%======================================================
A.K.G. and A.S. acknowledge SAMKHYA: High-Performance Computing Facility provided by Institute of Physics, Bhubaneswar, for numerical computations. TN acknowledges the NFSG ``NFSG/HYD/2023/H0911" from BITS Pilani.

\appendix
\newcounter{defcounter}
\setcounter{defcounter}{0}
%-------------------------------------------------
\section{Spatial symmetries and localization of corner states for asymmetric $\gamma$}{\label{App:A}
%-------------------------------------------------
%-------------------------------------------------
\begin{figure}[]
    \centering
    \includegraphics[width=0.99\linewidth]{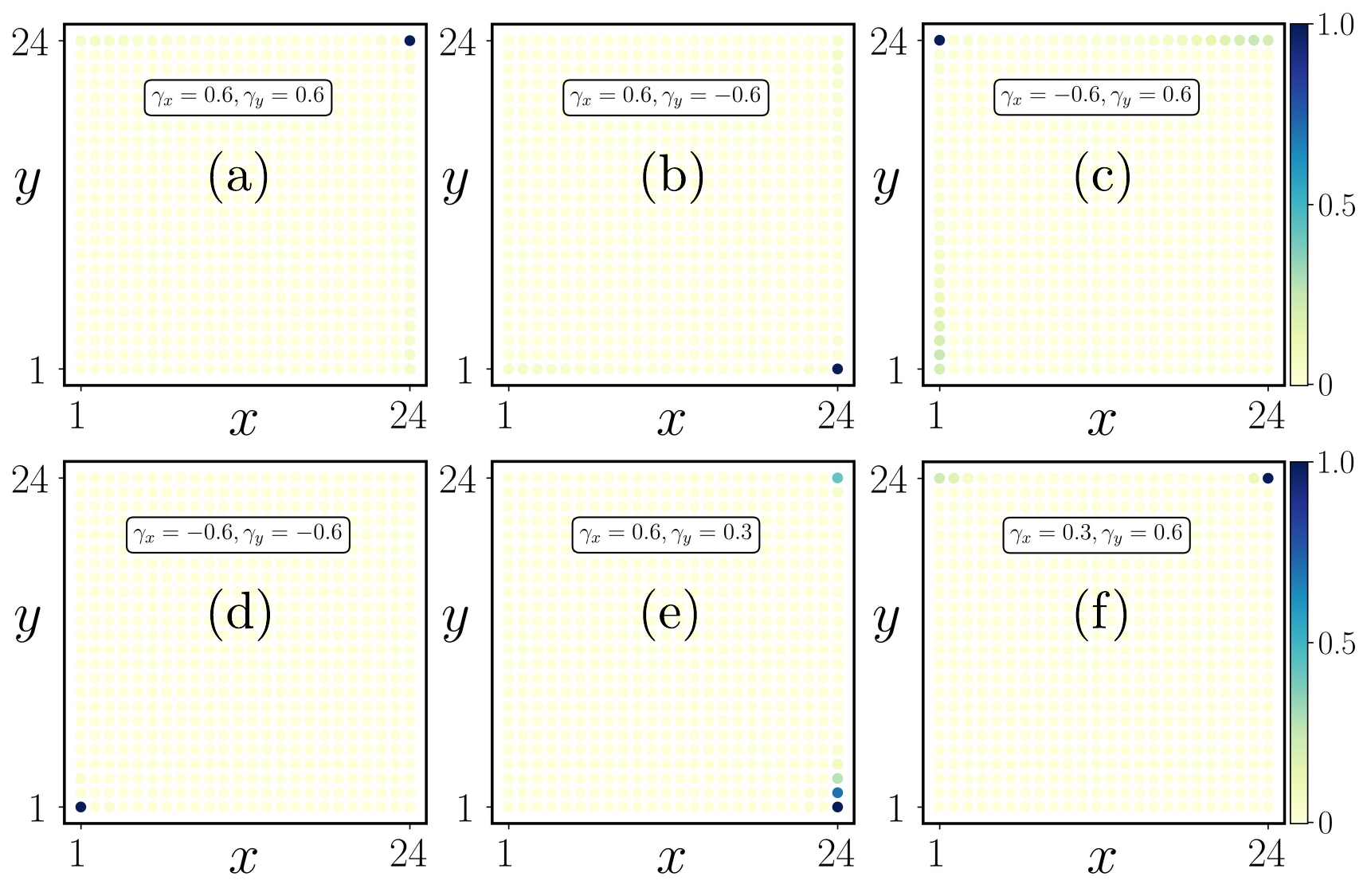}
    \caption{In 2D domain, we illustrate the LDOS spectrum associated with the $E=0$ states choosing different values of $\gamma_x$ and $\gamma_y$: (a) $\gamma_x=\gamma_y=0.6$, (b) $\gamma_x=-\gamma_y=0.6$, (c) $\gamma_x=-\gamma_y=-0.6$, (d) $\gamma_x=\gamma_y=-0.6$, (e) $\gamma_x=0.6,\gamma_y=0.3$, and (e) $\gamma_x=0.3,\gamma_y=0.6$. We choose $m_0=1.0$, while the other model parameters remain the same as mentioned in Fig.~\ref{fig:NHEigenNN}.}
    \label{fig:mirrorsymmetryLDOS}
\end{figure}
%-------------------------------------------------
%-------------------------------------------------
In Table~\ref{Table:symmetries}, we list all the symmetries that the model Hamiltonian $H_\gamma(\vect{k})$ breaks or preserves. As discussed in the main text, the mirror rotation symmetry $M_{xy}$ plays a crucial role in the localization of the corner states. In Fig.~\ref{fig:mirrorsymmetryLDOS}, we demonstrate the LDOS associated with the $E=0$ states choosing different values of $\gamma_x$ and $\gamma_y$. Note that, in the main text, we always consider 
$\gamma_x=\gamma_y=\gamma$. In Figs.~\ref{fig:mirrorsymmetryLDOS}(a-d), we illustrate the case when $\gamma_x$ and $\gamma_y$ carry the same amplitude but can have different signs for a NH
first NN Hamiltonian model. We observe that depending upon the signs of $\gamma_x$ and $\gamma_y$, the corner modes occupy different corners of the system. In contrast, when $\gamma_x\neq\gamma_y$, i.e., the mirror rotation symmetries are broken, the corner states can occupy more than one corner [see Figs.~\ref{fig:mirrorsymmetryLDOS}(e,f)]. However, it is to be noted that the localization 
at different corners carry different weights. %do not appear to be equal. 
}
%---------------------------------------------------------
%---------------------------------------------------------
\begin{figure}[htb]
    \centering
    \includegraphics[width=1.0\linewidth]{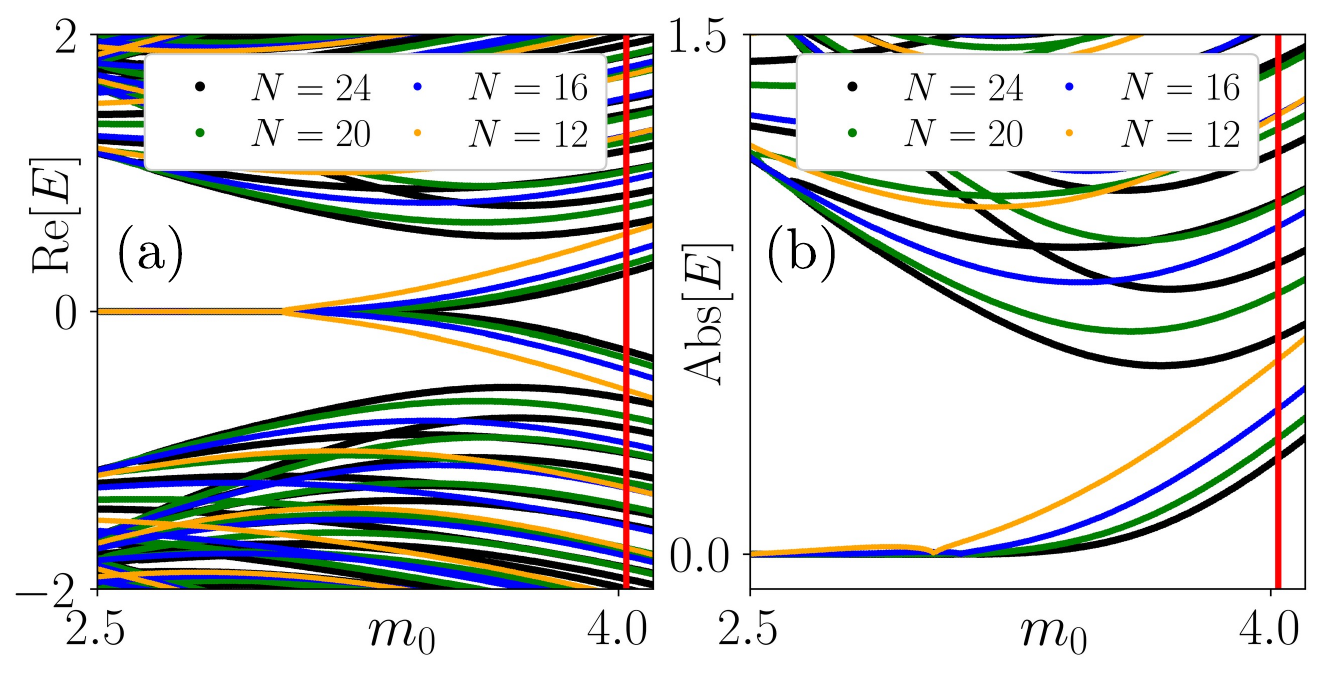} 
    \caption{In panels (a) and (b), we demonstrate ${\rm Re}[E]$ and ${\rm Abs}[E]$ close to the rightmost phase transition line in Figs.~\ref{fig:NHEigenNNN}(d,f), choosing different system sizes respectively.}
    \label{fig:finitesizescaling}
\end{figure}
%--------------------------------------------------------
%---------------------------------------------------------
%\textcolor{red}{
\section{Finite size scaling with second-NN NH hopping model}{\label{App:B}}
%----------------------------------------------------------
%----------------------------------------------------------
As discussed in the main text, the rightmost part of the phase boundary (in Figs.~\ref{fig:NHEigenNNN}(d,f)) corresponding to the second NN NH Hamiltonian encounters finite-size scaling. Here, we explicitly exhibit variation of the eigenvalue spectra considering different system sizes. In particular, we demonstrate the real (${\rm Re}[E]$) and absolute part (${\rm Abs}[E]$) of the eigenvalue spectra 
close to the rightmost phase transition line of Figs.~\ref{fig:NHEigenNNN}(d,f) choosing different system sizes ($N=12,16,20,24$) in Fig.~\ref{fig:finitesizescaling}. One can evidently observe from 
Figs.~\ref{fig:finitesizescaling}(a,b) that as we increase the system size, the zero-energy states in the eigenvalue spectra move towards the analytically obtained phase transition line (red). However, the translation towards the phase transition line as a function of the system size appears to be slower in nature.
%}

\bibliographystyle{apsrev4-2mod}
\bibliography{bibfile.bib}

%apsrev4-2.bst 2015-08-30 from 4.21a (PWD, AO, DPC/HNN) hacked
%Control: key (0)
%Control: author (72) initials jnrlst
%Control: editor formatted (1) identically to author
%Control: production of article title (-1) disabled
%Control: page (0) single
%Control: year (1) truncated
%Control: production of eprint (0) enabled
\begin{thebibliography}{93}%
\makeatletter
\providecommand \@ifxundefined [1]{%
 \@ifx{#1\undefined}
}%
\providecommand \@ifnum [1]{%
 \ifnum #1\expandafter \@firstoftwo
 \else \expandafter \@secondoftwo
 \fi
}%
\providecommand \@ifx [1]{%
 \ifx #1\expandafter \@firstoftwo
 \else \expandafter \@secondoftwo
 \fi
}%
\providecommand \natexlab [1]{#1}%
\providecommand \enquote  [1]{``#1''}%
\providecommand \bibnamefont  [1]{#1}%
\providecommand \bibfnamefont [1]{#1}%
\providecommand \citenamefont [1]{#1}%
\providecommand \href@noop [0]{\@secondoftwo}%
\providecommand \href [0]{\begingroup \@sanitize@url \@href}%
\providecommand \@href[1]{\@@startlink{#1}\@@href}%
\providecommand \@@href[1]{\endgroup#1\@@endlink}%
\providecommand \@sanitize@url [0]{\catcode `\\12\catcode `\$12\catcode
  `\&12\catcode `\#12\catcode `\^12\catcode `\_12\catcode `\%12\relax}%
\providecommand \@@startlink[1]{}%
\providecommand \@@endlink[0]{}%
\providecommand \url  [0]{\begingroup\@sanitize@url \@url }%
\providecommand \@url [1]{\endgroup\@href {#1}{\urlprefix }}%
\providecommand \urlprefix  [0]{URL }%
\providecommand \Eprint [0]{\href }%
\providecommand \doibase [0]{http://dx.doi.org/}%
\providecommand \selectlanguage [0]{\@gobble}%
\providecommand \bibinfo  [0]{\@secondoftwo}%
\providecommand \bibfield  [0]{\@secondoftwo}%
\providecommand \translation [1]{[#1]}%
\providecommand \BibitemOpen [0]{}%
\providecommand \bibitemStop [0]{}%
\providecommand \bibitemNoStop [0]{.\EOS\space}%
\providecommand \EOS [0]{\spacefactor3000\relax}%
\providecommand \BibitemShut  [1]{\csname bibitem#1\endcsname}%
\let\auto@bib@innerbib\@empty
%</preamble>
\bibitem [{\citenamefont {Hasan}\ and\ \citenamefont
  {Kane}(2010)}]{hasan2010colloquium}%
  \BibitemOpen
  \bibfield  {author} {\bibinfo {author} {\bibfnamefont {M.~Z.}\ \bibnamefont
  {Hasan}}\ and\ \bibinfo {author} {\bibfnamefont {C.~L.}\ \bibnamefont
  {Kane}},\ }\bibfield  {title} {\emph {\enquote {\bibinfo {title} {Colloquium:
  topological insulators},}\ }}\href {\doibase 10.1103/RevModPhys.82.3045}
  {\bibfield  {journal} {\bibinfo  {journal} {Rev. Mod. Phys.}\ }\textbf
  {\bibinfo {volume} {82}},\ \bibinfo {pages} {3045} (\bibinfo {year}
  {2010})}\BibitemShut {NoStop}%
\bibitem [{\citenamefont {Sato}\ and\ \citenamefont
  {Ando}(2017)}]{sato2017topological}%
  \BibitemOpen
  \bibfield  {author} {\bibinfo {author} {\bibfnamefont {M.}~\bibnamefont
  {Sato}}\ and\ \bibinfo {author} {\bibfnamefont {Y.}~\bibnamefont {Ando}},\
  }\bibfield  {title} {\emph {\enquote {\bibinfo {title} {Topological
  superconductors: a review},}\ }}\href@noop {} {\bibfield  {journal} {\bibinfo
   {journal} {Reports on Progress in Physics}\ }\textbf {\bibinfo {volume}
  {80}},\ \bibinfo {pages} {076501} (\bibinfo {year} {2017})}\BibitemShut
  {NoStop}%
\bibitem [{\citenamefont {Haldane}(1988)}]{Haldane88}%
  \BibitemOpen
  \bibfield  {author} {\bibinfo {author} {\bibfnamefont {F.~D.~M.}\
  \bibnamefont {Haldane}},\ }\bibfield  {title} {\emph {\enquote {\bibinfo
  {title} {Model for a Quantum Hall Effect without Landau Levels:
  Condensed-Matter Realization of the``Parity Anomaly"},}\ }}\href {\doibase
  10.1103/PhysRevLett.61.2015} {\bibfield  {journal} {\bibinfo  {journal}
  {Phys. Rev. Lett.}\ }\textbf {\bibinfo {volume} {61}},\ \bibinfo {pages}
  {2015} (\bibinfo {year} {1988})}\BibitemShut {NoStop}%
\bibitem [{\citenamefont {Kane}\ and\ \citenamefont {Mele}(2005)}]{Kane05}%
  \BibitemOpen
  \bibfield  {author} {\bibinfo {author} {\bibfnamefont {C.~L.}\ \bibnamefont
  {Kane}}\ and\ \bibinfo {author} {\bibfnamefont {E.~J.}\ \bibnamefont
  {Mele}},\ }\bibfield  {title} {\emph {\enquote {\bibinfo {title} {${Z}_{2}$
  Topological Order and the Quantum Spin Hall Effect},}\ }}\href {\doibase
  10.1103/PhysRevLett.95.146802} {\bibfield  {journal} {\bibinfo  {journal}
  {Phys. Rev. Lett.}\ }\textbf {\bibinfo {volume} {95}},\ \bibinfo {pages}
  {146802} (\bibinfo {year} {2005})}\BibitemShut {NoStop}%
\bibitem [{\citenamefont {Bernevig}\ \emph {et~al.}(2006)\citenamefont
  {Bernevig}, \citenamefont {Hughes},\ and\ \citenamefont
  {Zhang}}]{bernevig2006quantum}%
  \BibitemOpen
  \bibfield  {author} {\bibinfo {author} {\bibfnamefont {B.~A.}\ \bibnamefont
  {Bernevig}}, \bibinfo {author} {\bibfnamefont {T.~L.}\ \bibnamefont
  {Hughes}}, \ and\ \bibinfo {author} {\bibfnamefont {S.-C.}\ \bibnamefont
  {Zhang}},\ }\bibfield  {title} {\emph {\enquote {\bibinfo {title} {Quantum
  spin Hall effect and topological phase transition in HgTe quantum wells},}\
  }}\href {\doibase https://doi.org/10.1126/science.1133734} {\bibfield
  {journal} {\bibinfo  {journal} {Science}\ }\textbf {\bibinfo {volume}
  {314}},\ \bibinfo {pages} {1757} (\bibinfo {year} {2006})}\BibitemShut
  {NoStop}%
\bibitem [{\citenamefont {Benalcazar}\ \emph
  {et~al.}(2017{\natexlab{a}})\citenamefont {Benalcazar}, \citenamefont
  {Bernevig},\ and\ \citenamefont {Hughes}}]{benalcazar2017}%
  \BibitemOpen
  \bibfield  {author} {\bibinfo {author} {\bibfnamefont {W.~A.}\ \bibnamefont
  {Benalcazar}}, \bibinfo {author} {\bibfnamefont {B.~A.}\ \bibnamefont
  {Bernevig}}, \ and\ \bibinfo {author} {\bibfnamefont {T.~L.}\ \bibnamefont
  {Hughes}},\ }\bibfield  {title} {\emph {\enquote {\bibinfo {title} {Quantized
  electric multipole insulators},}\ }}\href {\doibase
  https://doi.org/10.1126/science.aah6442} {\bibfield  {journal} {\bibinfo
  {journal} {Science}\ }\textbf {\bibinfo {volume} {357}},\ \bibinfo {pages}
  {61} (\bibinfo {year} {2017}{\natexlab{a}})}\BibitemShut {NoStop}%
\bibitem [{\citenamefont {Benalcazar}\ \emph
  {et~al.}(2017{\natexlab{b}})\citenamefont {Benalcazar}, \citenamefont
  {Bernevig},\ and\ \citenamefont {Hughes}}]{benalcazarprb2017}%
  \BibitemOpen
  \bibfield  {author} {\bibinfo {author} {\bibfnamefont {W.~A.}\ \bibnamefont
  {Benalcazar}}, \bibinfo {author} {\bibfnamefont {B.~A.}\ \bibnamefont
  {Bernevig}}, \ and\ \bibinfo {author} {\bibfnamefont {T.~L.}\ \bibnamefont
  {Hughes}},\ }\bibfield  {title} {\emph {\enquote {\bibinfo {title} {Electric
  multipole moments, topological multipole moment pumping, and chiral hinge
  states in crystalline insulators},}\ }}\href {\doibase
  10.1103/PhysRevB.96.245115} {\bibfield  {journal} {\bibinfo  {journal} {Phys.
  Rev. B}\ }\textbf {\bibinfo {volume} {96}},\ \bibinfo {pages} {245115}
  (\bibinfo {year} {2017}{\natexlab{b}})}\BibitemShut {NoStop}%
\bibitem [{\citenamefont {Song}\ \emph {et~al.}(2017)\citenamefont {Song},
  \citenamefont {Fang},\ and\ \citenamefont {Fang}}]{Song2017}%
  \BibitemOpen
  \bibfield  {author} {\bibinfo {author} {\bibfnamefont {Z.}~\bibnamefont
  {Song}}, \bibinfo {author} {\bibfnamefont {Z.}~\bibnamefont {Fang}}, \ and\
  \bibinfo {author} {\bibfnamefont {C.}~\bibnamefont {Fang}},\ }\bibfield
  {title} {\emph {\enquote {\bibinfo {title} {$(d\ensuremath{-}2)$-Dimensional
  Edge States of Rotation Symmetry Protected Topological States},}\ }}\href
  {\doibase 10.1103/PhysRevLett.119.246402} {\bibfield  {journal} {\bibinfo
  {journal} {Phys. Rev. Lett.}\ }\textbf {\bibinfo {volume} {119}},\ \bibinfo
  {pages} {246402} (\bibinfo {year} {2017})}\BibitemShut {NoStop}%
\bibitem [{\citenamefont {Langbehn}\ \emph {et~al.}(2017)\citenamefont
  {Langbehn}, \citenamefont {Peng}, \citenamefont {Trifunovic}, \citenamefont
  {von Oppen},\ and\ \citenamefont {Brouwer}}]{Langbehn2017}%
  \BibitemOpen
  \bibfield  {author} {\bibinfo {author} {\bibfnamefont {J.}~\bibnamefont
  {Langbehn}}, \bibinfo {author} {\bibfnamefont {Y.}~\bibnamefont {Peng}},
  \bibinfo {author} {\bibfnamefont {L.}~\bibnamefont {Trifunovic}}, \bibinfo
  {author} {\bibfnamefont {F.}~\bibnamefont {von Oppen}}, \ and\ \bibinfo
  {author} {\bibfnamefont {P.~W.}\ \bibnamefont {Brouwer}},\ }\bibfield
  {title} {\emph {\enquote {\bibinfo {title} {Reflection-Symmetric Second-Order
  Topological Insulators and Superconductors},}\ }}\href {\doibase
  10.1103/PhysRevLett.119.246401} {\bibfield  {journal} {\bibinfo  {journal}
  {Phys. Rev. Lett.}\ }\textbf {\bibinfo {volume} {119}},\ \bibinfo {pages}
  {246401} (\bibinfo {year} {2017})}\BibitemShut {NoStop}%
\bibitem [{\citenamefont {Schindler}\ \emph {et~al.}(2018)\citenamefont
  {Schindler}, \citenamefont {Cook}, \citenamefont {Vergniory}, \citenamefont
  {Wang}, \citenamefont {Parkin}, \citenamefont {Bernevig},\ and\ \citenamefont
  {Neupert}}]{schindler2018}%
  \BibitemOpen
  \bibfield  {author} {\bibinfo {author} {\bibfnamefont {F.}~\bibnamefont
  {Schindler}}, \bibinfo {author} {\bibfnamefont {A.~M.}\ \bibnamefont {Cook}},
  \bibinfo {author} {\bibfnamefont {M.~G.}\ \bibnamefont {Vergniory}}, \bibinfo
  {author} {\bibfnamefont {Z.}~\bibnamefont {Wang}}, \bibinfo {author}
  {\bibfnamefont {S.~S.}\ \bibnamefont {Parkin}}, \bibinfo {author}
  {\bibfnamefont {B.~A.}\ \bibnamefont {Bernevig}}, \ and\ \bibinfo {author}
  {\bibfnamefont {T.}~\bibnamefont {Neupert}},\ }\bibfield  {title} {\emph
  {\enquote {\bibinfo {title} {Higher-order topological insulators},}\ }}\href
  {\doibase https://doi.org/10.1126/sciadv.aat0346} {\bibfield  {journal}
  {\bibinfo  {journal} {Science adv.}\ }\textbf {\bibinfo {volume} {4}},\
  \bibinfo {pages} {eaat0346} (\bibinfo {year} {2018})}\BibitemShut {NoStop}%
\bibitem [{\citenamefont {Franca}\ \emph {et~al.}(2018)\citenamefont {Franca},
  \citenamefont {van~den Brink},\ and\ \citenamefont {Fulga}}]{Franca2018}%
  \BibitemOpen
  \bibfield  {author} {\bibinfo {author} {\bibfnamefont {S.}~\bibnamefont
  {Franca}}, \bibinfo {author} {\bibfnamefont {J.}~\bibnamefont {van~den
  Brink}}, \ and\ \bibinfo {author} {\bibfnamefont {I.~C.}\ \bibnamefont
  {Fulga}},\ }\bibfield  {title} {\emph {\enquote {\bibinfo {title} {An
  anomalous higher-order topological insulator},}\ }}\href {\doibase
  10.1103/PhysRevB.98.201114} {\bibfield  {journal} {\bibinfo  {journal} {Phys.
  Rev. B}\ }\textbf {\bibinfo {volume} {98}},\ \bibinfo {pages} {201114}
  (\bibinfo {year} {2018})}\BibitemShut {NoStop}%
\bibitem [{\citenamefont {Wang}\ \emph {et~al.}(2019)\citenamefont {Wang},
  \citenamefont {Wieder}, \citenamefont {Li}, \citenamefont {Yan},\ and\
  \citenamefont {Bernevig}}]{wang2018higher}%
  \BibitemOpen
  \bibfield  {author} {\bibinfo {author} {\bibfnamefont {Z.}~\bibnamefont
  {Wang}}, \bibinfo {author} {\bibfnamefont {B.~J.}\ \bibnamefont {Wieder}},
  \bibinfo {author} {\bibfnamefont {J.}~\bibnamefont {Li}}, \bibinfo {author}
  {\bibfnamefont {B.}~\bibnamefont {Yan}}, \ and\ \bibinfo {author}
  {\bibfnamefont {B.~A.}\ \bibnamefont {Bernevig}},\ }\bibfield  {title} {\emph
  {\enquote {\bibinfo {title} {Higher-Order Topology, Monopole Nodal Lines, and
  the Origin of Large Fermi Arcs in Transition Metal Dichalcogenides
  $X{\mathrm{Te}}_{2}$ ($X=\mathrm{Mo},\mathrm{W}$)},}\ }}\href {\doibase
  10.1103/PhysRevLett.123.186401} {\bibfield  {journal} {\bibinfo  {journal}
  {Phys. Rev. Lett.}\ }\textbf {\bibinfo {volume} {123}},\ \bibinfo {pages}
  {186401} (\bibinfo {year} {2019})}\BibitemShut {NoStop}%
\bibitem [{\citenamefont {C\ifmmode \u{a}\else \u{a}\fi{}lug\ifmmode~\u{a}\else
  \u{a}\fi{}ru}\ \emph {et~al.}(2019)\citenamefont {C\ifmmode \u{a}\else
  \u{a}\fi{}lug\ifmmode~\u{a}\else \u{a}\fi{}ru}, \citenamefont {Juri\ifmmode
  \check{c}\else \v{c}\fi{}i\ifmmode~\acute{c}\else \'{c}\fi{}},\ and\
  \citenamefont {Roy}}]{Roy2019}%
  \BibitemOpen
  \bibfield  {author} {\bibinfo {author} {\bibfnamefont {D.}~\bibnamefont
  {C\ifmmode \u{a}\else \u{a}\fi{}lug\ifmmode~\u{a}\else \u{a}\fi{}ru}},
  \bibinfo {author} {\bibfnamefont {V.}~\bibnamefont {Juri\ifmmode
  \check{c}\else \v{c}\fi{}i\ifmmode~\acute{c}\else \'{c}\fi{}}}, \ and\
  \bibinfo {author} {\bibfnamefont {B.}~\bibnamefont {Roy}},\ }\bibfield
  {title} {\emph {\enquote {\bibinfo {title} {Higher-order topological phases:
  A general principle of construction},}\ }}\href {\doibase
  10.1103/PhysRevB.99.041301} {\bibfield  {journal} {\bibinfo  {journal} {Phys.
  Rev. B}\ }\textbf {\bibinfo {volume} {99}},\ \bibinfo {pages} {041301}
  (\bibinfo {year} {2019})}\BibitemShut {NoStop}%
\bibitem [{\citenamefont {Szumniak}\ \emph {et~al.}(2020)\citenamefont
  {Szumniak}, \citenamefont {Loss},\ and\ \citenamefont
  {Klinovaja}}]{Szumniak2020}%
  \BibitemOpen
  \bibfield  {author} {\bibinfo {author} {\bibfnamefont {P.}~\bibnamefont
  {Szumniak}}, \bibinfo {author} {\bibfnamefont {D.}~\bibnamefont {Loss}}, \
  and\ \bibinfo {author} {\bibfnamefont {J.}~\bibnamefont {Klinovaja}},\
  }\bibfield  {title} {\emph {\enquote {\bibinfo {title} {Hinge modes and
  surface states in second-order topological three-dimensional quantum Hall
  systems induced by charge density modulation},}\ }}\href {\doibase
  10.1103/PhysRevB.102.125126} {\bibfield  {journal} {\bibinfo  {journal}
  {Phys. Rev. B}\ }\textbf {\bibinfo {volume} {102}},\ \bibinfo {pages}
  {125126} (\bibinfo {year} {2020})}\BibitemShut {NoStop}%
\bibitem [{\citenamefont {Ni}\ \emph {et~al.}(2020)\citenamefont {Ni},
  \citenamefont {Li}, \citenamefont {Weiner}, \citenamefont {Al{\`u}},\ and\
  \citenamefont {Khanikaev}}]{Ni2020}%
  \BibitemOpen
  \bibfield  {author} {\bibinfo {author} {\bibfnamefont {X.}~\bibnamefont
  {Ni}}, \bibinfo {author} {\bibfnamefont {M.}~\bibnamefont {Li}}, \bibinfo
  {author} {\bibfnamefont {M.}~\bibnamefont {Weiner}}, \bibinfo {author}
  {\bibfnamefont {A.}~\bibnamefont {Al{\`u}}}, \ and\ \bibinfo {author}
  {\bibfnamefont {A.~B.}\ \bibnamefont {Khanikaev}},\ }\bibfield  {title}
  {\emph {\enquote {\bibinfo {title} {Demonstration of a quantized acoustic
  octupole topological insulator},}\ }}\href {\doibase
  10.1038/s41467-020-15705-y} {\bibfield  {journal} {\bibinfo  {journal}
  {Nature Communications}\ }\textbf {\bibinfo {volume} {11}},\ \bibinfo {pages}
  {2108} (\bibinfo {year} {2020})}\BibitemShut {NoStop}%
\bibitem [{\citenamefont {Xie}\ \emph {et~al.}(2021)\citenamefont {Xie},
  \citenamefont {Wang}, \citenamefont {Zhang}, \citenamefont {Zhan},
  \citenamefont {Jiang}, \citenamefont {Lu},\ and\ \citenamefont
  {Chen}}]{BiyeXie2021}%
  \BibitemOpen
  \bibfield  {author} {\bibinfo {author} {\bibfnamefont {B.}~\bibnamefont
  {Xie}}, \bibinfo {author} {\bibfnamefont {H.}~\bibnamefont {Wang}}, \bibinfo
  {author} {\bibfnamefont {X.}~\bibnamefont {Zhang}}, \bibinfo {author}
  {\bibfnamefont {P.}~\bibnamefont {Zhan}}, \bibinfo {author} {\bibfnamefont
  {J.}~\bibnamefont {Jiang}}, \bibinfo {author} {\bibfnamefont
  {M.}~\bibnamefont {Lu}}, \ and\ \bibinfo {author} {\bibfnamefont
  {Y.}~\bibnamefont {Chen}},\ }\bibfield  {title} {\emph {\enquote {\bibinfo
  {title} {Higher-order band topology},}\ }}\href {\doibase
  https://doi.org/10.1038/s42254-021-00323-4} {\bibfield  {journal} {\bibinfo
  {journal} {Nat. Rev. Phys.}\ }\textbf {\bibinfo {volume} {3}},\ \bibinfo
  {pages} {520} (\bibinfo {year} {2021})}\BibitemShut {NoStop}%
\bibitem [{\citenamefont {Saha}\ \emph {et~al.}(2022)\citenamefont {Saha},
  \citenamefont {Nag},\ and\ \citenamefont {Mandal}}]{saha2022dipolar}%
  \BibitemOpen
  \bibfield  {author} {\bibinfo {author} {\bibfnamefont {S.}~\bibnamefont
  {Saha}}, \bibinfo {author} {\bibfnamefont {T.}~\bibnamefont {Nag}}, \ and\
  \bibinfo {author} {\bibfnamefont {S.}~\bibnamefont {Mandal}},\ }\href@noop {}
  {\enquote {\bibinfo {title} {Dipolar quantum spin Hall insulator phase in
  extended Haldane model},}\ } (\bibinfo {year} {2022}),\ \Eprint
  {http://arxiv.org/abs/2204.06641}{arXiv:2204.06641}\BibitemShut {NoStop}%
\bibitem [{\citenamefont {Zhu}(2018)}]{Zhu2018}%
  \BibitemOpen
  \bibfield  {author} {\bibinfo {author} {\bibfnamefont {X.}~\bibnamefont
  {Zhu}},\ }\bibfield  {title} {\emph {\enquote {\bibinfo {title} {Tunable
  Majorana corner states in a two-dimensional second-order topological
  superconductor induced by magnetic fields},}\ }}\href {\doibase
  10.1103/PhysRevB.97.205134} {\bibfield  {journal} {\bibinfo  {journal} {Phys.
  Rev. B}\ }\textbf {\bibinfo {volume} {97}},\ \bibinfo {pages} {205134}
  (\bibinfo {year} {2018})}\BibitemShut {NoStop}%
\bibitem [{\citenamefont {Liu}\ \emph {et~al.}(2018)\citenamefont {Liu},
  \citenamefont {He},\ and\ \citenamefont {Nori}}]{Liu2018}%
  \BibitemOpen
  \bibfield  {author} {\bibinfo {author} {\bibfnamefont {T.}~\bibnamefont
  {Liu}}, \bibinfo {author} {\bibfnamefont {J.~J.}\ \bibnamefont {He}}, \ and\
  \bibinfo {author} {\bibfnamefont {F.}~\bibnamefont {Nori}},\ }\bibfield
  {title} {\emph {\enquote {\bibinfo {title} {Majorana corner states in a
  two-dimensional magnetic topological insulator on a high-temperature
  superconductor},}\ }}\href {\doibase 10.1103/PhysRevB.98.245413} {\bibfield
  {journal} {\bibinfo  {journal} {Phys. Rev. B}\ }\textbf {\bibinfo {volume}
  {98}},\ \bibinfo {pages} {245413} (\bibinfo {year} {2018})}\BibitemShut
  {NoStop}%
\bibitem [{\citenamefont {Yan}\ \emph {et~al.}(2018)\citenamefont {Yan},
  \citenamefont {Song},\ and\ \citenamefont {Wang}}]{Yan2018}%
  \BibitemOpen
  \bibfield  {author} {\bibinfo {author} {\bibfnamefont {Z.}~\bibnamefont
  {Yan}}, \bibinfo {author} {\bibfnamefont {F.}~\bibnamefont {Song}}, \ and\
  \bibinfo {author} {\bibfnamefont {Z.}~\bibnamefont {Wang}},\ }\bibfield
  {title} {\emph {\enquote {\bibinfo {title} {Majorana Corner Modes in a
  High-Temperature Platform},}\ }}\href {\doibase
  10.1103/PhysRevLett.121.096803} {\bibfield  {journal} {\bibinfo  {journal}
  {Phys. Rev. Lett.}\ }\textbf {\bibinfo {volume} {121}},\ \bibinfo {pages}
  {096803} (\bibinfo {year} {2018})}\BibitemShut {NoStop}%
\bibitem [{\citenamefont {Wang}\ \emph
  {et~al.}(2018{\natexlab{a}})\citenamefont {Wang}, \citenamefont {Lin},\ and\
  \citenamefont {Hughes}}]{WangWeak2018}%
  \BibitemOpen
  \bibfield  {author} {\bibinfo {author} {\bibfnamefont {Y.}~\bibnamefont
  {Wang}}, \bibinfo {author} {\bibfnamefont {M.}~\bibnamefont {Lin}}, \ and\
  \bibinfo {author} {\bibfnamefont {T.~L.}\ \bibnamefont {Hughes}},\ }\bibfield
   {title} {\emph {\enquote {\bibinfo {title} {Weak-pairing higher order
  topological superconductors},}\ }}\href {\doibase 10.1103/PhysRevB.98.165144}
  {\bibfield  {journal} {\bibinfo  {journal} {Phys. Rev. B}\ }\textbf {\bibinfo
  {volume} {98}},\ \bibinfo {pages} {165144} (\bibinfo {year}
  {2018}{\natexlab{a}})}\BibitemShut {NoStop}%
\bibitem [{\citenamefont {Zhang}\ \emph
  {et~al.}(2019{\natexlab{a}})\citenamefont {Zhang}, \citenamefont {Cole},\
  and\ \citenamefont {Das~Sarma}}]{Zhang2019}%
  \BibitemOpen
  \bibfield  {author} {\bibinfo {author} {\bibfnamefont {R.-X.}\ \bibnamefont
  {Zhang}}, \bibinfo {author} {\bibfnamefont {W.~S.}\ \bibnamefont {Cole}}, \
  and\ \bibinfo {author} {\bibfnamefont {S.}~\bibnamefont {Das~Sarma}},\
  }\bibfield  {title} {\emph {\enquote {\bibinfo {title} {Helical Hinge
  Majorana Modes in Iron-Based Superconductors},}\ }}\href {\doibase
  10.1103/PhysRevLett.122.187001} {\bibfield  {journal} {\bibinfo  {journal}
  {Phys. Rev. Lett.}\ }\textbf {\bibinfo {volume} {122}},\ \bibinfo {pages}
  {187001} (\bibinfo {year} {2019}{\natexlab{a}})}\BibitemShut {NoStop}%
\bibitem [{\citenamefont {Zhang}\ \emph
  {et~al.}(2019{\natexlab{b}})\citenamefont {Zhang}, \citenamefont {Cole},
  \citenamefont {Wu},\ and\ \citenamefont {Das~Sarma}}]{ZhangFe2019PRL}%
  \BibitemOpen
  \bibfield  {author} {\bibinfo {author} {\bibfnamefont {R.-X.}\ \bibnamefont
  {Zhang}}, \bibinfo {author} {\bibfnamefont {W.~S.}\ \bibnamefont {Cole}},
  \bibinfo {author} {\bibfnamefont {X.}~\bibnamefont {Wu}}, \ and\ \bibinfo
  {author} {\bibfnamefont {S.}~\bibnamefont {Das~Sarma}},\ }\bibfield  {title}
  {\emph {\enquote {\bibinfo {title} {Higher-Order Topology and Nodal
  Topological Superconductivity in Fe(Se,Te) Heterostructures},}\ }}\href
  {\doibase 10.1103/PhysRevLett.123.167001} {\bibfield  {journal} {\bibinfo
  {journal} {Phys. Rev. Lett.}\ }\textbf {\bibinfo {volume} {123}},\ \bibinfo
  {pages} {167001} (\bibinfo {year} {2019}{\natexlab{b}})}\BibitemShut
  {NoStop}%
\bibitem [{\citenamefont {Volpez}\ \emph {et~al.}(2019)\citenamefont {Volpez},
  \citenamefont {Loss},\ and\ \citenamefont {Klinovaja}}]{Volpez2019}%
  \BibitemOpen
  \bibfield  {author} {\bibinfo {author} {\bibfnamefont {Y.}~\bibnamefont
  {Volpez}}, \bibinfo {author} {\bibfnamefont {D.}~\bibnamefont {Loss}}, \ and\
  \bibinfo {author} {\bibfnamefont {J.}~\bibnamefont {Klinovaja}},\ }\bibfield
  {title} {\emph {\enquote {\bibinfo {title} {Second-Order Topological
  Superconductivity in $\ensuremath{\pi}$-Junction Rashba Layers},}\ }}\href
  {\doibase 10.1103/PhysRevLett.122.126402} {\bibfield  {journal} {\bibinfo
  {journal} {Phys. Rev. Lett.}\ }\textbf {\bibinfo {volume} {122}},\ \bibinfo
  {pages} {126402} (\bibinfo {year} {2019})}\BibitemShut {NoStop}%
\bibitem [{\citenamefont {Yan}(2019{\natexlab{a}})}]{YanPRB2019}%
  \BibitemOpen
  \bibfield  {author} {\bibinfo {author} {\bibfnamefont {Z.}~\bibnamefont
  {Yan}},\ }\bibfield  {title} {\emph {\enquote {\bibinfo {title} {Majorana
  corner and hinge modes in second-order topological insulator/superconductor
  heterostructures},}\ }}\href {\doibase 10.1103/PhysRevB.100.205406}
  {\bibfield  {journal} {\bibinfo  {journal} {Phys. Rev. B}\ }\textbf {\bibinfo
  {volume} {100}},\ \bibinfo {pages} {205406} (\bibinfo {year}
  {2019}{\natexlab{a}})}\BibitemShut {NoStop}%
\bibitem [{\citenamefont {Ghorashi}\ \emph {et~al.}(2019)\citenamefont
  {Ghorashi}, \citenamefont {Hu}, \citenamefont {Hughes},\ and\ \citenamefont
  {Rossi}}]{Ghorashi2019}%
  \BibitemOpen
  \bibfield  {author} {\bibinfo {author} {\bibfnamefont {S.~A.~A.}\
  \bibnamefont {Ghorashi}}, \bibinfo {author} {\bibfnamefont {X.}~\bibnamefont
  {Hu}}, \bibinfo {author} {\bibfnamefont {T.~L.}\ \bibnamefont {Hughes}}, \
  and\ \bibinfo {author} {\bibfnamefont {E.}~\bibnamefont {Rossi}},\ }\bibfield
   {title} {\emph {\enquote {\bibinfo {title} {Second-order Dirac
  superconductors and magnetic field induced Majorana hinge modes},}\ }}\href
  {\doibase 10.1103/PhysRevB.100.020509} {\bibfield  {journal} {\bibinfo
  {journal} {Phys. Rev. B}\ }\textbf {\bibinfo {volume} {100}},\ \bibinfo
  {pages} {020509} (\bibinfo {year} {2019})}\BibitemShut {NoStop}%
\bibitem [{\citenamefont {Wu}\ \emph {et~al.}(2020)\citenamefont {Wu},
  \citenamefont {Hou}, \citenamefont {Li}, \citenamefont {Luo}, \citenamefont
  {Shi},\ and\ \citenamefont {Zhang}}]{Wu2020}%
  \BibitemOpen
  \bibfield  {author} {\bibinfo {author} {\bibfnamefont {Y.-J.}\ \bibnamefont
  {Wu}}, \bibinfo {author} {\bibfnamefont {J.}~\bibnamefont {Hou}}, \bibinfo
  {author} {\bibfnamefont {Y.-M.}\ \bibnamefont {Li}}, \bibinfo {author}
  {\bibfnamefont {X.-W.}\ \bibnamefont {Luo}}, \bibinfo {author} {\bibfnamefont
  {X.}~\bibnamefont {Shi}}, \ and\ \bibinfo {author} {\bibfnamefont
  {C.}~\bibnamefont {Zhang}},\ }\bibfield  {title} {\emph {\enquote {\bibinfo
  {title} {In-Plane Zeeman-Field-Induced Majorana Corner and Hinge Modes in an
  $s$-Wave Superconductor Heterostructure},}\ }}\href {\doibase
  10.1103/PhysRevLett.124.227001} {\bibfield  {journal} {\bibinfo  {journal}
  {Phys. Rev. Lett.}\ }\textbf {\bibinfo {volume} {124}},\ \bibinfo {pages}
  {227001} (\bibinfo {year} {2020})}\BibitemShut {NoStop}%
\bibitem [{\citenamefont {Laubscher}\ \emph {et~al.}(2020)\citenamefont
  {Laubscher}, \citenamefont {Chughtai}, \citenamefont {Loss},\ and\
  \citenamefont {Klinovaja}}]{jelena2020HOTSC}%
  \BibitemOpen
  \bibfield  {author} {\bibinfo {author} {\bibfnamefont {K.}~\bibnamefont
  {Laubscher}}, \bibinfo {author} {\bibfnamefont {D.}~\bibnamefont {Chughtai}},
  \bibinfo {author} {\bibfnamefont {D.}~\bibnamefont {Loss}}, \ and\ \bibinfo
  {author} {\bibfnamefont {J.}~\bibnamefont {Klinovaja}},\ }\bibfield  {title}
  {\emph {\enquote {\bibinfo {title} {Kramers pairs of Majorana corner states
  in a topological insulator bilayer},}\ }}\href {\doibase
  10.1103/PhysRevB.102.195401} {\bibfield  {journal} {\bibinfo  {journal}
  {Phys. Rev. B}\ }\textbf {\bibinfo {volume} {102}},\ \bibinfo {pages}
  {195401} (\bibinfo {year} {2020})}\BibitemShut {NoStop}%
\bibitem [{\citenamefont {Roy}(2020)}]{BitanTSC2020}%
  \BibitemOpen
  \bibfield  {author} {\bibinfo {author} {\bibfnamefont {B.}~\bibnamefont
  {Roy}},\ }\bibfield  {title} {\emph {\enquote {\bibinfo {title} {Higher-order
  topological superconductors in $\mathcal{P}$-, $\mathcal{T}$-odd quadrupolar
  Dirac materials},}\ }}\href {\doibase 10.1103/PhysRevB.101.220506} {\bibfield
   {journal} {\bibinfo  {journal} {Phys. Rev. B}\ }\textbf {\bibinfo {volume}
  {101}},\ \bibinfo {pages} {220506} (\bibinfo {year} {2020})}\BibitemShut
  {NoStop}%
\bibitem [{\citenamefont {Zhang}\ \emph {et~al.}(2020)\citenamefont {Zhang},
  \citenamefont {Rui}, \citenamefont {Calzona}, \citenamefont {Choi},
  \citenamefont {Schnyder},\ and\ \citenamefont {Trauzettel}}]{SongboPRR22020}%
  \BibitemOpen
  \bibfield  {author} {\bibinfo {author} {\bibfnamefont {S.-B.}\ \bibnamefont
  {Zhang}}, \bibinfo {author} {\bibfnamefont {W.~B.}\ \bibnamefont {Rui}},
  \bibinfo {author} {\bibfnamefont {A.}~\bibnamefont {Calzona}}, \bibinfo
  {author} {\bibfnamefont {S.-J.}\ \bibnamefont {Choi}}, \bibinfo {author}
  {\bibfnamefont {A.~P.}\ \bibnamefont {Schnyder}}, \ and\ \bibinfo {author}
  {\bibfnamefont {B.}~\bibnamefont {Trauzettel}},\ }\bibfield  {title} {\emph
  {\enquote {\bibinfo {title} {Topological and holonomic quantum computation
  based on second-order topological superconductors},}\ }}\href {\doibase
  10.1103/PhysRevResearch.2.043025} {\bibfield  {journal} {\bibinfo  {journal}
  {Phys. Rev. Research}\ }\textbf {\bibinfo {volume} {2}},\ \bibinfo {pages}
  {043025} (\bibinfo {year} {2020})}\BibitemShut {NoStop}%
\bibitem [{\citenamefont {Kheirkhah}\ \emph {et~al.}(2021)\citenamefont
  {Kheirkhah}, \citenamefont {Yan},\ and\ \citenamefont
  {Marsiglio}}]{kheirkhah2020vortex}%
  \BibitemOpen
  \bibfield  {author} {\bibinfo {author} {\bibfnamefont {M.}~\bibnamefont
  {Kheirkhah}}, \bibinfo {author} {\bibfnamefont {Z.}~\bibnamefont {Yan}}, \
  and\ \bibinfo {author} {\bibfnamefont {F.}~\bibnamefont {Marsiglio}},\
  }\bibfield  {title} {\emph {\enquote {\bibinfo {title} {Vortex-line topology
  in iron-based superconductors with and without second-order topology},}\
  }}\href {\doibase 10.1103/PhysRevB.103.L140502} {\bibfield  {journal}
  {\bibinfo  {journal} {Phys. Rev. B}\ }\textbf {\bibinfo {volume} {103}},\
  \bibinfo {pages} {L140502} (\bibinfo {year} {2021})}\BibitemShut {NoStop}%
\bibitem [{\citenamefont {Yan}(2019{\natexlab{b}})}]{YanPRL2019}%
  \BibitemOpen
  \bibfield  {author} {\bibinfo {author} {\bibfnamefont {Z.}~\bibnamefont
  {Yan}},\ }\bibfield  {title} {\emph {\enquote {\bibinfo {title} {Higher-Order
  Topological Odd-Parity Superconductors},}\ }}\href {\doibase
  10.1103/PhysRevLett.123.177001} {\bibfield  {journal} {\bibinfo  {journal}
  {Phys. Rev. Lett.}\ }\textbf {\bibinfo {volume} {123}},\ \bibinfo {pages}
  {177001} (\bibinfo {year} {2019}{\natexlab{b}})}\BibitemShut {NoStop}%
\bibitem [{\citenamefont {Ahn}\ and\ \citenamefont {Yang}(2020)}]{AhnPRL2020}%
  \BibitemOpen
  \bibfield  {author} {\bibinfo {author} {\bibfnamefont {J.}~\bibnamefont
  {Ahn}}\ and\ \bibinfo {author} {\bibfnamefont {B.-J.}\ \bibnamefont {Yang}},\
  }\bibfield  {title} {\emph {\enquote {\bibinfo {title} {Higher-order
  topological superconductivity of spin-polarized fermions},}\ }}\href
  {\doibase 10.1103/PhysRevResearch.2.012060} {\bibfield  {journal} {\bibinfo
  {journal} {Phys. Rev. Research}\ }\textbf {\bibinfo {volume} {2}},\ \bibinfo
  {pages} {012060} (\bibinfo {year} {2020})}\BibitemShut {NoStop}%
\bibitem [{\citenamefont {Luo}\ \emph {et~al.}(2021)\citenamefont {Luo},
  \citenamefont {Pan},\ and\ \citenamefont {Liu}}]{luo2021higherorder2021}%
  \BibitemOpen
  \bibfield  {author} {\bibinfo {author} {\bibfnamefont {X.-J.}\ \bibnamefont
  {Luo}}, \bibinfo {author} {\bibfnamefont {X.-H.}\ \bibnamefont {Pan}}, \ and\
  \bibinfo {author} {\bibfnamefont {X.}~\bibnamefont {Liu}},\ }\bibfield
  {title} {\emph {\enquote {\bibinfo {title} {Higher-order topological
  superconductors based on weak topological insulators},}\ }}\href {\doibase
  10.1103/PhysRevB.104.104510} {\bibfield  {journal} {\bibinfo  {journal}
  {Phys. Rev. B}\ }\textbf {\bibinfo {volume} {104}},\ \bibinfo {pages}
  {104510} (\bibinfo {year} {2021})}\BibitemShut {NoStop}%
\bibitem [{\citenamefont {Wang}\ \emph
  {et~al.}(2018{\natexlab{b}})\citenamefont {Wang}, \citenamefont {Liu},
  \citenamefont {Lu},\ and\ \citenamefont {Zhang}}]{QWang2018}%
  \BibitemOpen
  \bibfield  {author} {\bibinfo {author} {\bibfnamefont {Q.}~\bibnamefont
  {Wang}}, \bibinfo {author} {\bibfnamefont {C.-C.}\ \bibnamefont {Liu}},
  \bibinfo {author} {\bibfnamefont {Y.-M.}\ \bibnamefont {Lu}}, \ and\ \bibinfo
  {author} {\bibfnamefont {F.}~\bibnamefont {Zhang}},\ }\bibfield  {title}
  {\emph {\enquote {\bibinfo {title} {High-Temperature Majorana Corner
  States},}\ }}\href {\doibase 10.1103/PhysRevLett.121.186801} {\bibfield
  {journal} {\bibinfo  {journal} {Phys. Rev. Lett.}\ }\textbf {\bibinfo
  {volume} {121}},\ \bibinfo {pages} {186801} (\bibinfo {year}
  {2018}{\natexlab{b}})}\BibitemShut {NoStop}%
\bibitem [{\citenamefont {Ghosh}\ \emph {et~al.}(2021)\citenamefont {Ghosh},
  \citenamefont {Nag},\ and\ \citenamefont {Saha}}]{Ghosh2021PRB}%
  \BibitemOpen
  \bibfield  {author} {\bibinfo {author} {\bibfnamefont {A.~K.}\ \bibnamefont
  {Ghosh}}, \bibinfo {author} {\bibfnamefont {T.}~\bibnamefont {Nag}}, \ and\
  \bibinfo {author} {\bibfnamefont {A.}~\bibnamefont {Saha}},\ }\bibfield
  {title} {\emph {\enquote {\bibinfo {title} {Hierarchy of higher-order
  topological superconductors in three dimensions},}\ }}\href {\doibase
  10.1103/PhysRevB.104.134508} {\bibfield  {journal} {\bibinfo  {journal}
  {Phys. Rev. B}\ }\textbf {\bibinfo {volume} {104}},\ \bibinfo {pages}
  {134508} (\bibinfo {year} {2021})}\BibitemShut {NoStop}%
\bibitem [{\citenamefont {Roy}\ and\ \citenamefont {Juri\ifmmode \check{c}\else
  \v{c}\fi{}i\ifmmode~\acute{c}\else \'{c}\fi{}}(2021)}]{RoyPRBL2021}%
  \BibitemOpen
  \bibfield  {author} {\bibinfo {author} {\bibfnamefont {B.}~\bibnamefont
  {Roy}}\ and\ \bibinfo {author} {\bibfnamefont {V.}~\bibnamefont {Juri\ifmmode
  \check{c}\else \v{c}\fi{}i\ifmmode~\acute{c}\else \'{c}\fi{}}},\ }\bibfield
  {title} {\emph {\enquote {\bibinfo {title} {Mixed-parity octupolar pairing
  and corner Majorana modes in three dimensions},}\ }}\href {\doibase
  10.1103/PhysRevB.104.L180503} {\bibfield  {journal} {\bibinfo  {journal}
  {Phys. Rev. B}\ }\textbf {\bibinfo {volume} {104}},\ \bibinfo {pages}
  {L180503} (\bibinfo {year} {2021})}\BibitemShut {NoStop}%
\bibitem [{\citenamefont {Li}\ \emph {et~al.}(2021)\citenamefont {Li},
  \citenamefont {Geier}, \citenamefont {Ingham},\ and\ \citenamefont
  {Scammell}}]{li2021higher}%
  \BibitemOpen
  \bibfield  {author} {\bibinfo {author} {\bibfnamefont {T.}~\bibnamefont
  {Li}}, \bibinfo {author} {\bibfnamefont {M.}~\bibnamefont {Geier}}, \bibinfo
  {author} {\bibfnamefont {J.}~\bibnamefont {Ingham}}, \ and\ \bibinfo {author}
  {\bibfnamefont {H.~D.}\ \bibnamefont {Scammell}},\ }\bibfield  {title} {\emph
  {\enquote {\bibinfo {title} {Higher-order topological superconductivity from
  repulsive interactions in kagome and honeycomb systems},}\ }}\href@noop {}
  {\bibfield  {journal} {\bibinfo  {journal} {2D Materials}\ }\textbf {\bibinfo
  {volume} {9}},\ \bibinfo {pages} {015031} (\bibinfo {year}
  {2021})}\BibitemShut {NoStop}%
\bibitem [{\citenamefont {Ghosh}\ \emph
  {et~al.}(2022{\natexlab{a}})\citenamefont {Ghosh}, \citenamefont {Nag},\ and\
  \citenamefont {Saha}}]{ghosh2022systematic}%
  \BibitemOpen
  \bibfield  {author} {\bibinfo {author} {\bibfnamefont {A.~K.}\ \bibnamefont
  {Ghosh}}, \bibinfo {author} {\bibfnamefont {T.}~\bibnamefont {Nag}}, \ and\
  \bibinfo {author} {\bibfnamefont {A.}~\bibnamefont {Saha}},\ }\bibfield
  {title} {\emph {\enquote {\bibinfo {title} {Systematic generation of the
  cascade of anomalous dynamical first- and higher-order modes in Floquet
  topological insulators},}\ }}\href {\doibase 10.1103/PhysRevB.105.115418}
  {\bibfield  {journal} {\bibinfo  {journal} {Phys. Rev. B}\ }\textbf {\bibinfo
  {volume} {105}},\ \bibinfo {pages} {115418} (\bibinfo {year}
  {2022}{\natexlab{a}})}\BibitemShut {NoStop}%
\bibitem [{\citenamefont {Ghosh}\ \emph
  {et~al.}(2022{\natexlab{b}})\citenamefont {Ghosh}, \citenamefont {Nag},\ and\
  \citenamefont {Saha}}]{ghosh2022dynamical}%
  \BibitemOpen
  \bibfield  {author} {\bibinfo {author} {\bibfnamefont {A.~K.}\ \bibnamefont
  {Ghosh}}, \bibinfo {author} {\bibfnamefont {T.}~\bibnamefont {Nag}}, \ and\
  \bibinfo {author} {\bibfnamefont {A.}~\bibnamefont {Saha}},\ }\bibfield
  {title} {\emph {\enquote {\bibinfo {title} {Dynamical construction of
  quadrupolar and octupolar topological superconductors},}\ }}\href {\doibase
  10.1103/PhysRevB.105.155406} {\bibfield  {journal} {\bibinfo  {journal}
  {Phys. Rev. B}\ }\textbf {\bibinfo {volume} {105}},\ \bibinfo {pages}
  {155406} (\bibinfo {year} {2022}{\natexlab{b}})}\BibitemShut {NoStop}%
\bibitem [{\citenamefont {DeGottardi}\ \emph {et~al.}(2013)\citenamefont
  {DeGottardi}, \citenamefont {Thakurathi}, \citenamefont {Vishveshwara},\ and\
  \citenamefont {Sen}}]{DeGottardi13}%
  \BibitemOpen
  \bibfield  {author} {\bibinfo {author} {\bibfnamefont {W.}~\bibnamefont
  {DeGottardi}}, \bibinfo {author} {\bibfnamefont {M.}~\bibnamefont
  {Thakurathi}}, \bibinfo {author} {\bibfnamefont {S.}~\bibnamefont
  {Vishveshwara}}, \ and\ \bibinfo {author} {\bibfnamefont {D.}~\bibnamefont
  {Sen}},\ }\bibfield  {title} {\emph {\enquote {\bibinfo {title} {Majorana
  fermions in superconducting wires: Effects of long-range hopping, broken
  time-reversal symmetry, and potential landscapes},}\ }}\href {\doibase
  10.1103/PhysRevB.88.165111} {\bibfield  {journal} {\bibinfo  {journal} {Phys.
  Rev. B}\ }\textbf {\bibinfo {volume} {88}},\ \bibinfo {pages} {165111}
  (\bibinfo {year} {2013})}\BibitemShut {NoStop}%
\bibitem [{\citenamefont {Niu}\ \emph {et~al.}(2012)\citenamefont {Niu},
  \citenamefont {Chung}, \citenamefont {Hsu}, \citenamefont {Mandal},
  \citenamefont {Raghu},\ and\ \citenamefont {Chakravarty}}]{Niu12}%
  \BibitemOpen
  \bibfield  {author} {\bibinfo {author} {\bibfnamefont {Y.}~\bibnamefont
  {Niu}}, \bibinfo {author} {\bibfnamefont {S.~B.}\ \bibnamefont {Chung}},
  \bibinfo {author} {\bibfnamefont {C.-H.}\ \bibnamefont {Hsu}}, \bibinfo
  {author} {\bibfnamefont {I.}~\bibnamefont {Mandal}}, \bibinfo {author}
  {\bibfnamefont {S.}~\bibnamefont {Raghu}}, \ and\ \bibinfo {author}
  {\bibfnamefont {S.}~\bibnamefont {Chakravarty}},\ }\bibfield  {title} {\emph
  {\enquote {\bibinfo {title} {Majorana zero modes in a quantum Ising chain
  with longer-ranged interactions},}\ }}\href {\doibase
  10.1103/PhysRevB.85.035110} {\bibfield  {journal} {\bibinfo  {journal} {Phys.
  Rev. B}\ }\textbf {\bibinfo {volume} {85}},\ \bibinfo {pages} {035110}
  (\bibinfo {year} {2012})}\BibitemShut {NoStop}%
\bibitem [{\citenamefont {Xie}\ \emph {et~al.}(2019)\citenamefont {Xie},
  \citenamefont {Gou}, \citenamefont {Xiao}, \citenamefont {Gadway},\ and\
  \citenamefont {Yan}}]{xie2019topological}%
  \BibitemOpen
  \bibfield  {author} {\bibinfo {author} {\bibfnamefont {D.}~\bibnamefont
  {Xie}}, \bibinfo {author} {\bibfnamefont {W.}~\bibnamefont {Gou}}, \bibinfo
  {author} {\bibfnamefont {T.}~\bibnamefont {Xiao}}, \bibinfo {author}
  {\bibfnamefont {B.}~\bibnamefont {Gadway}}, \ and\ \bibinfo {author}
  {\bibfnamefont {B.}~\bibnamefont {Yan}},\ }\bibfield  {title} {\emph
  {\enquote {\bibinfo {title} {Topological characterizations of an extended
  Su--Schrieffer--Heeger model},}\ }}\href@noop {} {\bibfield  {journal}
  {\bibinfo  {journal} {npj Quantum Information}\ }\textbf {\bibinfo {volume}
  {5}},\ \bibinfo {pages} {55} (\bibinfo {year} {2019})}\BibitemShut {NoStop}%
\bibitem [{\citenamefont {Dias}\ and\ \citenamefont {Marques}(2022)}]{Dias22}%
  \BibitemOpen
  \bibfield  {author} {\bibinfo {author} {\bibfnamefont {R.~G.}\ \bibnamefont
  {Dias}}\ and\ \bibinfo {author} {\bibfnamefont {A.~M.}\ \bibnamefont
  {Marques}},\ }\bibfield  {title} {\emph {\enquote {\bibinfo {title}
  {Long-range hopping and indexing assumption in one-dimensional topological
  insulators},}\ }}\href {\doibase 10.1103/PhysRevB.105.035102} {\bibfield
  {journal} {\bibinfo  {journal} {Phys. Rev. B}\ }\textbf {\bibinfo {volume}
  {105}},\ \bibinfo {pages} {035102} (\bibinfo {year} {2022})}\BibitemShut
  {NoStop}%
\bibitem [{\citenamefont {Mondal}\ and\ \citenamefont
  {Nag}()}]{mondal2024persistent}%
  \BibitemOpen
  \bibfield  {author} {\bibinfo {author} {\bibfnamefont {D.}~\bibnamefont
  {Mondal}}\ and\ \bibinfo {author} {\bibfnamefont {T.}~\bibnamefont {Nag}},\
  }\href@noop {} {\enquote {\bibinfo {title} {Persistent anomaly in dynamical
  quantum phase transition in long-range non-Hermitian $p$-wave Kitaev
  chian},}\ }\Eprint
  {http://arxiv.org/abs/2402.04603}{arXiv:2402.04603}\BibitemShut {NoStop}%
\bibitem [{\citenamefont {Thakurathi}\ \emph {et~al.}(2013)\citenamefont
  {Thakurathi}, \citenamefont {Patel}, \citenamefont {Sen},\ and\ \citenamefont
  {Dutta}}]{Thakurathi2013}%
  \BibitemOpen
  \bibfield  {author} {\bibinfo {author} {\bibfnamefont {M.}~\bibnamefont
  {Thakurathi}}, \bibinfo {author} {\bibfnamefont {A.~A.}\ \bibnamefont
  {Patel}}, \bibinfo {author} {\bibfnamefont {D.}~\bibnamefont {Sen}}, \ and\
  \bibinfo {author} {\bibfnamefont {A.}~\bibnamefont {Dutta}},\ }\bibfield
  {title} {\emph {\enquote {\bibinfo {title} {Floquet generation of Majorana
  end modes and topological invariants},}\ }}\href {\doibase
  10.1103/PhysRevB.88.155133} {\bibfield  {journal} {\bibinfo  {journal} {Phys.
  Rev. B}\ }\textbf {\bibinfo {volume} {88}},\ \bibinfo {pages} {155133}
  (\bibinfo {year} {2013})}\BibitemShut {NoStop}%
\bibitem [{\citenamefont {Schnyder}\ \emph {et~al.}(2008)\citenamefont
  {Schnyder}, \citenamefont {Ryu}, \citenamefont {Furusaki},\ and\
  \citenamefont {Ludwig}}]{Schnyder08}%
  \BibitemOpen
  \bibfield  {author} {\bibinfo {author} {\bibfnamefont {A.~P.}\ \bibnamefont
  {Schnyder}}, \bibinfo {author} {\bibfnamefont {S.}~\bibnamefont {Ryu}},
  \bibinfo {author} {\bibfnamefont {A.}~\bibnamefont {Furusaki}}, \ and\
  \bibinfo {author} {\bibfnamefont {A.~W.~W.}\ \bibnamefont {Ludwig}},\
  }\bibfield  {title} {\emph {\enquote {\bibinfo {title} {Classification of
  topological insulators and superconductors in three spatial dimensions},}\
  }}\href {\doibase 10.1103/PhysRevB.78.195125} {\bibfield  {journal} {\bibinfo
   {journal} {Phys. Rev. B}\ }\textbf {\bibinfo {volume} {78}},\ \bibinfo
  {pages} {195125} (\bibinfo {year} {2008})}\BibitemShut {NoStop}%
\bibitem [{\citenamefont {Kitaev}(2009)}]{kitaev2009periodic}%
  \BibitemOpen
  \bibfield  {author} {\bibinfo {author} {\bibfnamefont {A.}~\bibnamefont
  {Kitaev}},\ }in\ \href@noop {} {\emph {\bibinfo {booktitle} {AIP conference
  proceedings}}},\ Vol.\ \bibinfo {volume} {1134}\ (\bibinfo {organization}
  {American Institute of Physics},\ \bibinfo {year} {2009})\ pp.\ \bibinfo
  {pages} {22--30}\BibitemShut {NoStop}%
\bibitem [{\citenamefont {Ryu}\ \emph {et~al.}(2010)\citenamefont {Ryu},
  \citenamefont {Schnyder}, \citenamefont {Furusaki},\ and\ \citenamefont
  {Ludwig}}]{ryu2010topological}%
  \BibitemOpen
  \bibfield  {author} {\bibinfo {author} {\bibfnamefont {S.}~\bibnamefont
  {Ryu}}, \bibinfo {author} {\bibfnamefont {A.~P.}\ \bibnamefont {Schnyder}},
  \bibinfo {author} {\bibfnamefont {A.}~\bibnamefont {Furusaki}}, \ and\
  \bibinfo {author} {\bibfnamefont {A.~W.}\ \bibnamefont {Ludwig}},\ }\bibfield
   {title} {\emph {\enquote {\bibinfo {title} {Topological insulators and
  superconductors: tenfold way and dimensional hierarchy},}\ }}\href@noop {}
  {\bibfield  {journal} {\bibinfo  {journal} {New Journal of Physics}\ }\textbf
  {\bibinfo {volume} {12}},\ \bibinfo {pages} {065010} (\bibinfo {year}
  {2010})}\BibitemShut {NoStop}%
\bibitem [{\citenamefont {Ryu}\ and\ \citenamefont {Hatsugai}(2002)}]{Ryu02}%
  \BibitemOpen
  \bibfield  {author} {\bibinfo {author} {\bibfnamefont {S.}~\bibnamefont
  {Ryu}}\ and\ \bibinfo {author} {\bibfnamefont {Y.}~\bibnamefont {Hatsugai}},\
  }\bibfield  {title} {\emph {\enquote {\bibinfo {title} {Topological Origin of
  Zero-Energy Edge States in Particle-Hole Symmetric Systems},}\ }}\href
  {\doibase 10.1103/PhysRevLett.89.077002} {\bibfield  {journal} {\bibinfo
  {journal} {Phys. Rev. Lett.}\ }\textbf {\bibinfo {volume} {89}},\ \bibinfo
  {pages} {077002} (\bibinfo {year} {2002})}\BibitemShut {NoStop}%
\bibitem [{\citenamefont {Teo}\ and\ \citenamefont {Kane}(2010)}]{Teo10}%
  \BibitemOpen
  \bibfield  {author} {\bibinfo {author} {\bibfnamefont {J.~C.~Y.}\
  \bibnamefont {Teo}}\ and\ \bibinfo {author} {\bibfnamefont {C.~L.}\
  \bibnamefont {Kane}},\ }\bibfield  {title} {\emph {\enquote {\bibinfo {title}
  {Topological defects and gapless modes in insulators and superconductors},}\
  }}\href {\doibase 10.1103/PhysRevB.82.115120} {\bibfield  {journal} {\bibinfo
   {journal} {Phys. Rev. B}\ }\textbf {\bibinfo {volume} {82}},\ \bibinfo
  {pages} {115120} (\bibinfo {year} {2010})}\BibitemShut {NoStop}%
\bibitem [{\citenamefont {Nakahara}(2018)}]{nakahara2018geometry}%
  \BibitemOpen
  \bibfield  {author} {\bibinfo {author} {\bibfnamefont {M.}~\bibnamefont
  {Nakahara}},\ }\href@noop {} {\emph {\bibinfo {title} {Geometry, topology and
  physics}}}\ (\bibinfo  {publisher} {CRC press},\ \bibinfo {year}
  {2018})\BibitemShut {NoStop}%
\bibitem [{\citenamefont {Benalcazar}\ and\ \citenamefont
  {Cerjan}(2022)}]{Benalcazar22}%
  \BibitemOpen
  \bibfield  {author} {\bibinfo {author} {\bibfnamefont {W.~A.}\ \bibnamefont
  {Benalcazar}}\ and\ \bibinfo {author} {\bibfnamefont {A.}~\bibnamefont
  {Cerjan}},\ }\bibfield  {title} {\emph {\enquote {\bibinfo {title}
  {Chiral-Symmetric Higher-Order Topological Phases of Matter},}\ }}\href
  {\doibase 10.1103/PhysRevLett.128.127601} {\bibfield  {journal} {\bibinfo
  {journal} {Phys. Rev. Lett.}\ }\textbf {\bibinfo {volume} {128}},\ \bibinfo
  {pages} {127601} (\bibinfo {year} {2022})}\BibitemShut {NoStop}%
\bibitem [{\citenamefont {Parappurath}\ \emph {et~al.}(2020)\citenamefont
  {Parappurath}, \citenamefont {Alpeggiani}, \citenamefont {Kuipers},\ and\
  \citenamefont {Verhagen}}]{parappurath2020direct}%
  \BibitemOpen
  \bibfield  {author} {\bibinfo {author} {\bibfnamefont {N.}~\bibnamefont
  {Parappurath}}, \bibinfo {author} {\bibfnamefont {F.}~\bibnamefont
  {Alpeggiani}}, \bibinfo {author} {\bibfnamefont {L.}~\bibnamefont {Kuipers}},
  \ and\ \bibinfo {author} {\bibfnamefont {E.}~\bibnamefont {Verhagen}},\
  }\bibfield  {title} {\emph {\enquote {\bibinfo {title} {Direct observation of
  topological edge states in silicon photonic crystals: Spin, dispersion, and
  chiral routing},}\ }}\href {\doibase 10.1126/sciadv.aaw4137} {\bibfield
  {journal} {\bibinfo  {journal} {Science Advances}\ }\textbf {\bibinfo
  {volume} {6}},\ \bibinfo {pages} {eaaw4137} (\bibinfo {year}
  {2020})}\BibitemShut {NoStop}%
\bibitem [{\citenamefont {Yang}\ \emph {et~al.}(2019)\citenamefont {Yang},
  \citenamefont {Gao}, \citenamefont {Xue}, \citenamefont {Zhang},
  \citenamefont {He}, \citenamefont {Yang}, \citenamefont {Singh},
  \citenamefont {Chong}, \citenamefont {Zhang},\ and\ \citenamefont
  {Chen}}]{yang2019realization}%
  \BibitemOpen
  \bibfield  {author} {\bibinfo {author} {\bibfnamefont {Y.}~\bibnamefont
  {Yang}}, \bibinfo {author} {\bibfnamefont {Z.}~\bibnamefont {Gao}}, \bibinfo
  {author} {\bibfnamefont {H.}~\bibnamefont {Xue}}, \bibinfo {author}
  {\bibfnamefont {L.}~\bibnamefont {Zhang}}, \bibinfo {author} {\bibfnamefont
  {M.}~\bibnamefont {He}}, \bibinfo {author} {\bibfnamefont {Z.}~\bibnamefont
  {Yang}}, \bibinfo {author} {\bibfnamefont {R.}~\bibnamefont {Singh}},
  \bibinfo {author} {\bibfnamefont {Y.}~\bibnamefont {Chong}}, \bibinfo
  {author} {\bibfnamefont {B.}~\bibnamefont {Zhang}}, \ and\ \bibinfo {author}
  {\bibfnamefont {H.}~\bibnamefont {Chen}},\ }\bibfield  {title} {\emph
  {\enquote {\bibinfo {title} {Realization of a three-dimensional photonic
  topological insulator},}\ }}\href {\doibase 10.1038/s41586-018-0829-0}
  {\bibfield  {journal} {\bibinfo  {journal} {Nature}\ }\textbf {\bibinfo
  {volume} {565}},\ \bibinfo {pages} {622} (\bibinfo {year}
  {2019})}\BibitemShut {NoStop}%
\bibitem [{\citenamefont {Malzard}\ \emph {et~al.}(2015)\citenamefont
  {Malzard}, \citenamefont {Poli},\ and\ \citenamefont
  {Schomerus}}]{Malzard15}%
  \BibitemOpen
  \bibfield  {author} {\bibinfo {author} {\bibfnamefont {S.}~\bibnamefont
  {Malzard}}, \bibinfo {author} {\bibfnamefont {C.}~\bibnamefont {Poli}}, \
  and\ \bibinfo {author} {\bibfnamefont {H.}~\bibnamefont {Schomerus}},\
  }\bibfield  {title} {\emph {\enquote {\bibinfo {title} {Topologically
  Protected Defect States in Open Photonic Systems with Non-Hermitian
  Charge-Conjugation and Parity-Time Symmetry},}\ }}\href {\doibase
  10.1103/PhysRevLett.115.200402} {\bibfield  {journal} {\bibinfo  {journal}
  {Phys. Rev. Lett.}\ }\textbf {\bibinfo {volume} {115}},\ \bibinfo {pages}
  {200402} (\bibinfo {year} {2015})}\BibitemShut {NoStop}%
\bibitem [{\citenamefont {Regensburger}\ \emph {et~al.}(2012)\citenamefont
  {Regensburger}, \citenamefont {Bersch}, \citenamefont {Miri}, \citenamefont
  {Onishchukov}, \citenamefont {Christodoulides},\ and\ \citenamefont
  {Peschel}}]{regensburger2012parity}%
  \BibitemOpen
  \bibfield  {author} {\bibinfo {author} {\bibfnamefont {A.}~\bibnamefont
  {Regensburger}}, \bibinfo {author} {\bibfnamefont {C.}~\bibnamefont
  {Bersch}}, \bibinfo {author} {\bibfnamefont {M.-A.}\ \bibnamefont {Miri}},
  \bibinfo {author} {\bibfnamefont {G.}~\bibnamefont {Onishchukov}}, \bibinfo
  {author} {\bibfnamefont {D.~N.}\ \bibnamefont {Christodoulides}}, \ and\
  \bibinfo {author} {\bibfnamefont {U.}~\bibnamefont {Peschel}},\ }\bibfield
  {title} {\emph {\enquote {\bibinfo {title} {Parity--time synthetic photonic
  lattices},}\ }}\href {\doibase 10.1038/nature11298} {\bibfield  {journal}
  {\bibinfo  {journal} {Nature}\ }\textbf {\bibinfo {volume} {488}},\ \bibinfo
  {pages} {167} (\bibinfo {year} {2012})}\BibitemShut {NoStop}%
\bibitem [{\citenamefont {El-Ganainy}\ \emph {et~al.}(2018)\citenamefont
  {El-Ganainy}, \citenamefont {Makris}, \citenamefont {Khajavikhan},
  \citenamefont {Musslimani}, \citenamefont {Rotter},\ and\ \citenamefont
  {Christodoulides}}]{el2018non}%
  \BibitemOpen
  \bibfield  {author} {\bibinfo {author} {\bibfnamefont {R.}~\bibnamefont
  {El-Ganainy}}, \bibinfo {author} {\bibfnamefont {K.~G.}\ \bibnamefont
  {Makris}}, \bibinfo {author} {\bibfnamefont {M.}~\bibnamefont {Khajavikhan}},
  \bibinfo {author} {\bibfnamefont {Z.~H.}\ \bibnamefont {Musslimani}},
  \bibinfo {author} {\bibfnamefont {S.}~\bibnamefont {Rotter}}, \ and\ \bibinfo
  {author} {\bibfnamefont {D.~N.}\ \bibnamefont {Christodoulides}},\ }\bibfield
   {title} {\emph {\enquote {\bibinfo {title} {Non-Hermitian physics and PT
  symmetry},}\ }}\href {\doibase 10.1038/nphys4323} {\bibfield  {journal}
  {\bibinfo  {journal} {Nature Physics}\ }\textbf {\bibinfo {volume} {14}},\
  \bibinfo {pages} {11} (\bibinfo {year} {2018})}\BibitemShut {NoStop}%
\bibitem [{\citenamefont {Denner}\ \emph {et~al.}(2021)\citenamefont {Denner},
  \citenamefont {Skurativska}, \citenamefont {Schindler}, \citenamefont
  {Fischer}, \citenamefont {Thomale}, \citenamefont {Bzdu{\v{s}}ek},\ and\
  \citenamefont {Neupert}}]{Denner2021}%
  \BibitemOpen
  \bibfield  {author} {\bibinfo {author} {\bibfnamefont {M.~M.}\ \bibnamefont
  {Denner}}, \bibinfo {author} {\bibfnamefont {A.}~\bibnamefont {Skurativska}},
  \bibinfo {author} {\bibfnamefont {F.}~\bibnamefont {Schindler}}, \bibinfo
  {author} {\bibfnamefont {M.~H.}\ \bibnamefont {Fischer}}, \bibinfo {author}
  {\bibfnamefont {R.}~\bibnamefont {Thomale}}, \bibinfo {author} {\bibfnamefont
  {T.}~\bibnamefont {Bzdu{\v{s}}ek}}, \ and\ \bibinfo {author} {\bibfnamefont
  {T.}~\bibnamefont {Neupert}},\ }\bibfield  {title} {\emph {\enquote {\bibinfo
  {title} {Exceptional topological insulators},}\ }}\href {\doibase
  10.1038/s41467-021-25947-z} {\bibfield  {journal} {\bibinfo  {journal}
  {Nature Communications}\ }\textbf {\bibinfo {volume} {12}},\ \bibinfo {pages}
  {5681} (\bibinfo {year} {2021})}\BibitemShut {NoStop}%
\bibitem [{\citenamefont {Bergholtz}\ and\ \citenamefont
  {Budich}(2019)}]{Bergholtz19}%
  \BibitemOpen
  \bibfield  {author} {\bibinfo {author} {\bibfnamefont {E.~J.}\ \bibnamefont
  {Bergholtz}}\ and\ \bibinfo {author} {\bibfnamefont {J.~C.}\ \bibnamefont
  {Budich}},\ }\bibfield  {title} {\emph {\enquote {\bibinfo {title}
  {Non-Hermitian Weyl physics in topological insulator ferromagnet
  junctions},}\ }}\href {\doibase 10.1103/PhysRevResearch.1.012003} {\bibfield
  {journal} {\bibinfo  {journal} {Phys. Rev. Research}\ }\textbf {\bibinfo
  {volume} {1}},\ \bibinfo {pages} {012003} (\bibinfo {year}
  {2019})}\BibitemShut {NoStop}%
\bibitem [{\citenamefont {Yang}\ \emph {et~al.}(2021)\citenamefont {Yang},
  \citenamefont {Li}, \citenamefont {Duan},\ and\ \citenamefont {Xu}}]{Yang21}%
  \BibitemOpen
  \bibfield  {author} {\bibinfo {author} {\bibfnamefont {Y.-B.}\ \bibnamefont
  {Yang}}, \bibinfo {author} {\bibfnamefont {K.}~\bibnamefont {Li}}, \bibinfo
  {author} {\bibfnamefont {L.-M.}\ \bibnamefont {Duan}}, \ and\ \bibinfo
  {author} {\bibfnamefont {Y.}~\bibnamefont {Xu}},\ }\bibfield  {title} {\emph
  {\enquote {\bibinfo {title} {Higher-order topological Anderson insulators},}\
  }}\href {\doibase 10.1103/PhysRevB.103.085408} {\bibfield  {journal}
  {\bibinfo  {journal} {Phys. Rev. B}\ }\textbf {\bibinfo {volume} {103}},\
  \bibinfo {pages} {085408} (\bibinfo {year} {2021})}\BibitemShut {NoStop}%
\bibitem [{\citenamefont {San-Jose}\ \emph {et~al.}(2016)\citenamefont
  {San-Jose}, \citenamefont {Cayao}, \citenamefont {Prada},\ and\ \citenamefont
  {Aguado}}]{San-Jose2016}%
  \BibitemOpen
  \bibfield  {author} {\bibinfo {author} {\bibfnamefont {P.}~\bibnamefont
  {San-Jose}}, \bibinfo {author} {\bibfnamefont {J.}~\bibnamefont {Cayao}},
  \bibinfo {author} {\bibfnamefont {E.}~\bibnamefont {Prada}}, \ and\ \bibinfo
  {author} {\bibfnamefont {R.}~\bibnamefont {Aguado}},\ }\bibfield  {title}
  {\emph {\enquote {\bibinfo {title} {Majorana bound states from exceptional
  points in non-topological superconductors},}\ }}\href {\doibase
  10.1038/srep21427} {\bibfield  {journal} {\bibinfo  {journal} {Scientific
  Reports}\ }\textbf {\bibinfo {volume} {6}},\ \bibinfo {pages} {21427}
  (\bibinfo {year} {2016})}\BibitemShut {NoStop}%
\bibitem [{\citenamefont {Kozii}\ and\ \citenamefont {Fu}()}]{kozii2017non}%
  \BibitemOpen
  \bibfield  {author} {\bibinfo {author} {\bibfnamefont {V.}~\bibnamefont
  {Kozii}}\ and\ \bibinfo {author} {\bibfnamefont {L.}~\bibnamefont {Fu}},\
  }\href@noop {} {\enquote {\bibinfo {title} {Non-Hermitian topological theory
  of finite-lifetime quasiparticles: prediction of bulk Fermi arc due to
  exceptional point},}\ }\Eprint
  {http://arxiv.org/abs/1708.05841}{arXiv:1708.05841}\BibitemShut {NoStop}%
\bibitem [{\citenamefont {Yoshida}\ \emph {et~al.}(2018)\citenamefont
  {Yoshida}, \citenamefont {Peters},\ and\ \citenamefont
  {Kawakami}}]{Yoshida18}%
  \BibitemOpen
  \bibfield  {author} {\bibinfo {author} {\bibfnamefont {T.}~\bibnamefont
  {Yoshida}}, \bibinfo {author} {\bibfnamefont {R.}~\bibnamefont {Peters}}, \
  and\ \bibinfo {author} {\bibfnamefont {N.}~\bibnamefont {Kawakami}},\
  }\bibfield  {title} {\emph {\enquote {\bibinfo {title} {Non-Hermitian
  perspective of the band structure in heavy-fermion systems},}\ }}\href
  {\doibase 10.1103/PhysRevB.98.035141} {\bibfield  {journal} {\bibinfo
  {journal} {Phys. Rev. B}\ }\textbf {\bibinfo {volume} {98}},\ \bibinfo
  {pages} {035141} (\bibinfo {year} {2018})}\BibitemShut {NoStop}%
\bibitem [{\citenamefont {Shen}\ and\ \citenamefont {Fu}(2018)}]{Shen18}%
  \BibitemOpen
  \bibfield  {author} {\bibinfo {author} {\bibfnamefont {H.}~\bibnamefont
  {Shen}}\ and\ \bibinfo {author} {\bibfnamefont {L.}~\bibnamefont {Fu}},\
  }\bibfield  {title} {\emph {\enquote {\bibinfo {title} {Quantum Oscillation
  from In-Gap States and a Non-Hermitian Landau Level Problem},}\ }}\href
  {\doibase 10.1103/PhysRevLett.121.026403} {\bibfield  {journal} {\bibinfo
  {journal} {Phys. Rev. Lett.}\ }\textbf {\bibinfo {volume} {121}},\ \bibinfo
  {pages} {026403} (\bibinfo {year} {2018})}\BibitemShut {NoStop}%
\bibitem [{\citenamefont {Musslimani}\ \emph {et~al.}(2008)\citenamefont
  {Musslimani}, \citenamefont {Makris}, \citenamefont {El-Ganainy},\ and\
  \citenamefont {Christodoulides}}]{Musslimani08}%
  \BibitemOpen
  \bibfield  {author} {\bibinfo {author} {\bibfnamefont {Z.~H.}\ \bibnamefont
  {Musslimani}}, \bibinfo {author} {\bibfnamefont {K.~G.}\ \bibnamefont
  {Makris}}, \bibinfo {author} {\bibfnamefont {R.}~\bibnamefont {El-Ganainy}},
  \ and\ \bibinfo {author} {\bibfnamefont {D.~N.}\ \bibnamefont
  {Christodoulides}},\ }\bibfield  {title} {\emph {\enquote {\bibinfo {title}
  {Optical Solitons in $\mathcal{P}\mathcal{T}$ Periodic Potentials},}\ }}\href
  {\doibase 10.1103/PhysRevLett.100.030402} {\bibfield  {journal} {\bibinfo
  {journal} {Phys. Rev. Lett.}\ }\textbf {\bibinfo {volume} {100}},\ \bibinfo
  {pages} {030402} (\bibinfo {year} {2008})}\BibitemShut {NoStop}%
\bibitem [{\citenamefont {Makris}\ \emph {et~al.}(2008)\citenamefont {Makris},
  \citenamefont {El-Ganainy}, \citenamefont {Christodoulides},\ and\
  \citenamefont {Musslimani}}]{Makris08}%
  \BibitemOpen
  \bibfield  {author} {\bibinfo {author} {\bibfnamefont {K.~G.}\ \bibnamefont
  {Makris}}, \bibinfo {author} {\bibfnamefont {R.}~\bibnamefont {El-Ganainy}},
  \bibinfo {author} {\bibfnamefont {D.~N.}\ \bibnamefont {Christodoulides}}, \
  and\ \bibinfo {author} {\bibfnamefont {Z.~H.}\ \bibnamefont {Musslimani}},\
  }\bibfield  {title} {\emph {\enquote {\bibinfo {title} {Beam Dynamics in
  $\mathcal{P}\mathcal{T}$ Symmetric Optical Lattices},}\ }}\href {\doibase
  10.1103/PhysRevLett.100.103904} {\bibfield  {journal} {\bibinfo  {journal}
  {Phys. Rev. Lett.}\ }\textbf {\bibinfo {volume} {100}},\ \bibinfo {pages}
  {103904} (\bibinfo {year} {2008})}\BibitemShut {NoStop}%
\bibitem [{\citenamefont {Yao}\ \emph {et~al.}(2018)\citenamefont {Yao},
  \citenamefont {Song},\ and\ \citenamefont {Wang}}]{YaoPRLSecond2018}%
  \BibitemOpen
  \bibfield  {author} {\bibinfo {author} {\bibfnamefont {S.}~\bibnamefont
  {Yao}}, \bibinfo {author} {\bibfnamefont {F.}~\bibnamefont {Song}}, \ and\
  \bibinfo {author} {\bibfnamefont {Z.}~\bibnamefont {Wang}},\ }\bibfield
  {title} {\emph {\enquote {\bibinfo {title} {Non-Hermitian Chern Bands},}\
  }}\href {\doibase 10.1103/PhysRevLett.121.136802} {\bibfield  {journal}
  {\bibinfo  {journal} {Phys. Rev. Lett.}\ }\textbf {\bibinfo {volume} {121}},\
  \bibinfo {pages} {136802} (\bibinfo {year} {2018})}\BibitemShut {NoStop}%
\bibitem [{\citenamefont {Kawabata}\ \emph {et~al.}(2019)\citenamefont
  {Kawabata}, \citenamefont {Shiozaki}, \citenamefont {Ueda},\ and\
  \citenamefont {Sato}}]{KawabataPRX2019}%
  \BibitemOpen
  \bibfield  {author} {\bibinfo {author} {\bibfnamefont {K.}~\bibnamefont
  {Kawabata}}, \bibinfo {author} {\bibfnamefont {K.}~\bibnamefont {Shiozaki}},
  \bibinfo {author} {\bibfnamefont {M.}~\bibnamefont {Ueda}}, \ and\ \bibinfo
  {author} {\bibfnamefont {M.}~\bibnamefont {Sato}},\ }\bibfield  {title}
  {\emph {\enquote {\bibinfo {title} {Symmetry and Topology in Non-Hermitian
  Physics},}\ }}\href {\doibase 10.1103/PhysRevX.9.041015} {\bibfield
  {journal} {\bibinfo  {journal} {Phys. Rev. X}\ }\textbf {\bibinfo {volume}
  {9}},\ \bibinfo {pages} {041015} (\bibinfo {year} {2019})}\BibitemShut
  {NoStop}%
\bibitem [{\citenamefont {Bergholtz}\ \emph {et~al.}(2021)\citenamefont
  {Bergholtz}, \citenamefont {Budich},\ and\ \citenamefont
  {Kunst}}]{Bergholtz2021}%
  \BibitemOpen
  \bibfield  {author} {\bibinfo {author} {\bibfnamefont {E.~J.}\ \bibnamefont
  {Bergholtz}}, \bibinfo {author} {\bibfnamefont {J.~C.}\ \bibnamefont
  {Budich}}, \ and\ \bibinfo {author} {\bibfnamefont {F.~K.}\ \bibnamefont
  {Kunst}},\ }\bibfield  {title} {\emph {\enquote {\bibinfo {title}
  {Exceptional topology of non-Hermitian systems},}\ }}\href {\doibase
  10.1103/RevModPhys.93.015005} {\bibfield  {journal} {\bibinfo  {journal}
  {Rev. Mod. Phys.}\ }\textbf {\bibinfo {volume} {93}},\ \bibinfo {pages}
  {015005} (\bibinfo {year} {2021})}\BibitemShut {NoStop}%
\bibitem [{\citenamefont {Sone}\ \emph {et~al.}(2020)\citenamefont {Sone},
  \citenamefont {Ashida},\ and\ \citenamefont {Sagawa}}]{Sone2020}%
  \BibitemOpen
  \bibfield  {author} {\bibinfo {author} {\bibfnamefont {K.}~\bibnamefont
  {Sone}}, \bibinfo {author} {\bibfnamefont {Y.}~\bibnamefont {Ashida}}, \ and\
  \bibinfo {author} {\bibfnamefont {T.}~\bibnamefont {Sagawa}},\ }\bibfield
  {title} {\emph {\enquote {\bibinfo {title} {Exceptional non-Hermitian
  topological edge mode and its application to active matter},}\ }}\href
  {\doibase 10.1038/s41467-020-19488-0} {\bibfield  {journal} {\bibinfo
  {journal} {Nature Communications}\ }\textbf {\bibinfo {volume} {11}},\
  \bibinfo {pages} {5745} (\bibinfo {year} {2020})}\BibitemShut {NoStop}%
\bibitem [{\citenamefont {Ghosh}\ and\ \citenamefont
  {Black-Schaffer}(2024)}]{ghosh2024majorana}%
  \BibitemOpen
  \bibfield  {author} {\bibinfo {author} {\bibfnamefont {A.~K.}\ \bibnamefont
  {Ghosh}}\ and\ \bibinfo {author} {\bibfnamefont {A.~M.}\ \bibnamefont
  {Black-Schaffer}},\ }\href@noop {} {\enquote {\bibinfo {title} {Majorana
  zero-modes in a dissipative Rashba nanowire},}\ } (\bibinfo {year} {2024}),\
  \Eprint {http://arxiv.org/abs/2403.00419}{arXiv:2403.00419
  [cond-mat.mes-hall]}\BibitemShut {NoStop}%
\bibitem [{\citenamefont {Yao}\ and\ \citenamefont {Wang}(2018)}]{YaoPRL2018}%
  \BibitemOpen
  \bibfield  {author} {\bibinfo {author} {\bibfnamefont {S.}~\bibnamefont
  {Yao}}\ and\ \bibinfo {author} {\bibfnamefont {Z.}~\bibnamefont {Wang}},\
  }\bibfield  {title} {\emph {\enquote {\bibinfo {title} {Edge States and
  Topological Invariants of Non-Hermitian Systems},}\ }}\href {\doibase
  10.1103/PhysRevLett.121.086803} {\bibfield  {journal} {\bibinfo  {journal}
  {Phys. Rev. Lett.}\ }\textbf {\bibinfo {volume} {121}},\ \bibinfo {pages}
  {086803} (\bibinfo {year} {2018})}\BibitemShut {NoStop}%
\bibitem [{\citenamefont {Kunst}\ \emph {et~al.}(2018)\citenamefont {Kunst},
  \citenamefont {Edvardsson}, \citenamefont {Budich},\ and\ \citenamefont
  {Bergholtz}}]{Kunst18}%
  \BibitemOpen
  \bibfield  {author} {\bibinfo {author} {\bibfnamefont {F.~K.}\ \bibnamefont
  {Kunst}}, \bibinfo {author} {\bibfnamefont {E.}~\bibnamefont {Edvardsson}},
  \bibinfo {author} {\bibfnamefont {J.~C.}\ \bibnamefont {Budich}}, \ and\
  \bibinfo {author} {\bibfnamefont {E.~J.}\ \bibnamefont {Bergholtz}},\
  }\bibfield  {title} {\emph {\enquote {\bibinfo {title} {Biorthogonal
  Bulk-Boundary Correspondence in Non-Hermitian Systems},}\ }}\href {\doibase
  10.1103/PhysRevLett.121.026808} {\bibfield  {journal} {\bibinfo  {journal}
  {Phys. Rev. Lett.}\ }\textbf {\bibinfo {volume} {121}},\ \bibinfo {pages}
  {026808} (\bibinfo {year} {2018})}\BibitemShut {NoStop}%
\bibitem [{\citenamefont {Liu}\ \emph {et~al.}(2019)\citenamefont {Liu},
  \citenamefont {Zhang}, \citenamefont {Ai}, \citenamefont {Gong},
  \citenamefont {Kawabata}, \citenamefont {Ueda},\ and\ \citenamefont
  {Nori}}]{NoriNHPRL2019}%
  \BibitemOpen
  \bibfield  {author} {\bibinfo {author} {\bibfnamefont {T.}~\bibnamefont
  {Liu}}, \bibinfo {author} {\bibfnamefont {Y.-R.}\ \bibnamefont {Zhang}},
  \bibinfo {author} {\bibfnamefont {Q.}~\bibnamefont {Ai}}, \bibinfo {author}
  {\bibfnamefont {Z.}~\bibnamefont {Gong}}, \bibinfo {author} {\bibfnamefont
  {K.}~\bibnamefont {Kawabata}}, \bibinfo {author} {\bibfnamefont
  {M.}~\bibnamefont {Ueda}}, \ and\ \bibinfo {author} {\bibfnamefont
  {F.}~\bibnamefont {Nori}},\ }\bibfield  {title} {\emph {\enquote {\bibinfo
  {title} {Second-Order Topological Phases in Non-Hermitian Systems},}\ }}\href
  {\doibase 10.1103/PhysRevLett.122.076801} {\bibfield  {journal} {\bibinfo
  {journal} {Phys. Rev. Lett.}\ }\textbf {\bibinfo {volume} {122}},\ \bibinfo
  {pages} {076801} (\bibinfo {year} {2019})}\BibitemShut {NoStop}%
\bibitem [{\citenamefont {Helbig}\ \emph {et~al.}(2020)\citenamefont {Helbig},
  \citenamefont {Hofmann}, \citenamefont {Imhof}, \citenamefont {Abdelghany},
  \citenamefont {Kiessling}, \citenamefont {Molenkamp}, \citenamefont {Lee},
  \citenamefont {Szameit}, \citenamefont {Greiter},\ and\ \citenamefont
  {Thomale}}]{helbig2020generalized}%
  \BibitemOpen
  \bibfield  {author} {\bibinfo {author} {\bibfnamefont {T.}~\bibnamefont
  {Helbig}}, \bibinfo {author} {\bibfnamefont {T.}~\bibnamefont {Hofmann}},
  \bibinfo {author} {\bibfnamefont {S.}~\bibnamefont {Imhof}}, \bibinfo
  {author} {\bibfnamefont {M.}~\bibnamefont {Abdelghany}}, \bibinfo {author}
  {\bibfnamefont {T.}~\bibnamefont {Kiessling}}, \bibinfo {author}
  {\bibfnamefont {L.~W.}\ \bibnamefont {Molenkamp}}, \bibinfo {author}
  {\bibfnamefont {C.~H.}\ \bibnamefont {Lee}}, \bibinfo {author} {\bibfnamefont
  {A.}~\bibnamefont {Szameit}}, \bibinfo {author} {\bibfnamefont
  {M.}~\bibnamefont {Greiter}}, \ and\ \bibinfo {author} {\bibfnamefont
  {R.}~\bibnamefont {Thomale}},\ }\bibfield  {title} {\emph {\enquote {\bibinfo
  {title} {Generalized bulk--boundary correspondence in non-Hermitian
  topolectrical circuits},}\ }}\href {\doibase 10.1038/s41567-020-0922-9}
  {\bibfield  {journal} {\bibinfo  {journal} {Nature Physics}\ }\textbf
  {\bibinfo {volume} {16}},\ \bibinfo {pages} {747} (\bibinfo {year}
  {2020})}\BibitemShut {NoStop}%
\bibitem [{\citenamefont {Borgnia}\ \emph {et~al.}(2020)\citenamefont
  {Borgnia}, \citenamefont {Kruchkov},\ and\ \citenamefont
  {Slager}}]{Borgnia20}%
  \BibitemOpen
  \bibfield  {author} {\bibinfo {author} {\bibfnamefont {D.~S.}\ \bibnamefont
  {Borgnia}}, \bibinfo {author} {\bibfnamefont {A.~J.}\ \bibnamefont
  {Kruchkov}}, \ and\ \bibinfo {author} {\bibfnamefont {R.-J.}\ \bibnamefont
  {Slager}},\ }\bibfield  {title} {\emph {\enquote {\bibinfo {title}
  {Non-Hermitian Boundary Modes and Topology},}\ }}\href {\doibase
  10.1103/PhysRevLett.124.056802} {\bibfield  {journal} {\bibinfo  {journal}
  {Phys. Rev. Lett.}\ }\textbf {\bibinfo {volume} {124}},\ \bibinfo {pages}
  {056802} (\bibinfo {year} {2020})}\BibitemShut {NoStop}%
\bibitem [{\citenamefont {Koch}\ and\ \citenamefont {Budich}(2020)}]{Koch2020}%
  \BibitemOpen
  \bibfield  {author} {\bibinfo {author} {\bibfnamefont {R.}~\bibnamefont
  {Koch}}\ and\ \bibinfo {author} {\bibfnamefont {J.~C.}\ \bibnamefont
  {Budich}},\ }\bibfield  {title} {\emph {\enquote {\bibinfo {title}
  {Bulk-boundary correspondence in non-Hermitian systems: stability analysis
  for generalized boundary conditions},}\ }}\href {\doibase
  10.1140/epjd/e2020-100641-y} {\bibfield  {journal} {\bibinfo  {journal} {The
  European Physical Journal D}\ }\textbf {\bibinfo {volume} {74}},\ \bibinfo
  {pages} {70} (\bibinfo {year} {2020})}\BibitemShut {NoStop}%
\bibitem [{\citenamefont {Zirnstein}\ \emph {et~al.}(2021)\citenamefont
  {Zirnstein}, \citenamefont {Refael},\ and\ \citenamefont
  {Rosenow}}]{ZirnsteinPRL2021}%
  \BibitemOpen
  \bibfield  {author} {\bibinfo {author} {\bibfnamefont {H.-G.}\ \bibnamefont
  {Zirnstein}}, \bibinfo {author} {\bibfnamefont {G.}~\bibnamefont {Refael}}, \
  and\ \bibinfo {author} {\bibfnamefont {B.}~\bibnamefont {Rosenow}},\
  }\bibfield  {title} {\emph {\enquote {\bibinfo {title} {Bulk-Boundary
  Correspondence for Non-Hermitian Hamiltonians via Green Functions},}\ }}\href
  {\doibase 10.1103/PhysRevLett.126.216407} {\bibfield  {journal} {\bibinfo
  {journal} {Phys. Rev. Lett.}\ }\textbf {\bibinfo {volume} {126}},\ \bibinfo
  {pages} {216407} (\bibinfo {year} {2021})}\BibitemShut {NoStop}%
\bibitem [{\citenamefont {Takane}(2021)}]{TakaneJPS2021}%
  \BibitemOpen
  \bibfield  {author} {\bibinfo {author} {\bibfnamefont {Y.}~\bibnamefont
  {Takane}},\ }\bibfield  {title} {\emph {\enquote {\bibinfo {title}
  {Bulk–Boundary Correspondence in a Non-Hermitian Chern Insulator},}\
  }}\href {\doibase 10.7566/JPSJ.90.033704} {\bibfield  {journal} {\bibinfo
  {journal} {Journal of the Physical Society of Japan}\ }\textbf {\bibinfo
  {volume} {90}},\ \bibinfo {pages} {033704} (\bibinfo {year}
  {2021})}\BibitemShut {NoStop}%
\bibitem [{\citenamefont {Kawabata}\ \emph {et~al.}(2020)\citenamefont
  {Kawabata}, \citenamefont {Sato},\ and\ \citenamefont
  {Shiozaki}}]{Kawabata20b}%
  \BibitemOpen
  \bibfield  {author} {\bibinfo {author} {\bibfnamefont {K.}~\bibnamefont
  {Kawabata}}, \bibinfo {author} {\bibfnamefont {M.}~\bibnamefont {Sato}}, \
  and\ \bibinfo {author} {\bibfnamefont {K.}~\bibnamefont {Shiozaki}},\
  }\bibfield  {title} {\emph {\enquote {\bibinfo {title} {Higher-order
  non-Hermitian skin effect},}\ }}\href {\doibase 10.1103/PhysRevB.102.205118}
  {\bibfield  {journal} {\bibinfo  {journal} {Phys. Rev. B}\ }\textbf {\bibinfo
  {volume} {102}},\ \bibinfo {pages} {205118} (\bibinfo {year}
  {2020})}\BibitemShut {NoStop}%
\bibitem [{\citenamefont {Bender}(2007)}]{bender2007making}%
  \BibitemOpen
  \bibfield  {author} {\bibinfo {author} {\bibfnamefont {C.~M.}\ \bibnamefont
  {Bender}},\ }\bibfield  {title} {\emph {\enquote {\bibinfo {title} {Making
  sense of non-Hermitian Hamiltonians},}\ }}\href {\doibase
  10.1088/0034-4885/70/6/r03} {\bibfield  {journal} {\bibinfo  {journal} {Rep.
  Prog. Phys.}\ }\textbf {\bibinfo {volume} {70}},\ \bibinfo {pages} {947}
  (\bibinfo {year} {2007})}\BibitemShut {NoStop}%
\bibitem [{\citenamefont {Heiss}(2012)}]{heiss2012physics}%
  \BibitemOpen
  \bibfield  {author} {\bibinfo {author} {\bibfnamefont {W.~D.}\ \bibnamefont
  {Heiss}},\ }\bibfield  {title} {\emph {\enquote {\bibinfo {title} {The
  physics of exceptional points},}\ }}\href {\doibase
  10.1088/1751-8113/45/44/444016} {\bibfield  {journal} {\bibinfo  {journal}
  {J. Phys. A: Math. Theor.}\ }\textbf {\bibinfo {volume} {45}},\ \bibinfo
  {pages} {444016} (\bibinfo {year} {2012})}\BibitemShut {NoStop}%
\bibitem [{\citenamefont {P\'erez-Gonz\'alez}\ \emph
  {et~al.}(2019)\citenamefont {P\'erez-Gonz\'alez}, \citenamefont {Bello},
  \citenamefont {G\'omez-Le\'on},\ and\ \citenamefont {Platero}}]{Beatriz19}%
  \BibitemOpen
  \bibfield  {author} {\bibinfo {author} {\bibfnamefont {B.}~\bibnamefont
  {P\'erez-Gonz\'alez}}, \bibinfo {author} {\bibfnamefont {M.}~\bibnamefont
  {Bello}}, \bibinfo {author} {\bibfnamefont {A.}~\bibnamefont
  {G\'omez-Le\'on}}, \ and\ \bibinfo {author} {\bibfnamefont {G.}~\bibnamefont
  {Platero}},\ }\bibfield  {title} {\emph {\enquote {\bibinfo {title}
  {Interplay between long-range hopping and disorder in topological systems},}\
  }}\href {\doibase 10.1103/PhysRevB.99.035146} {\bibfield  {journal} {\bibinfo
   {journal} {Phys. Rev. B}\ }\textbf {\bibinfo {volume} {99}},\ \bibinfo
  {pages} {035146} (\bibinfo {year} {2019})}\BibitemShut {NoStop}%
\bibitem [{\citenamefont {An}\ \emph {et~al.}(2018)\citenamefont {An},
  \citenamefont {Meier},\ and\ \citenamefont {Gadway}}]{Fangzhao18}%
  \BibitemOpen
  \bibfield  {author} {\bibinfo {author} {\bibfnamefont {F.~A.}\ \bibnamefont
  {An}}, \bibinfo {author} {\bibfnamefont {E.~J.}\ \bibnamefont {Meier}}, \
  and\ \bibinfo {author} {\bibfnamefont {B.}~\bibnamefont {Gadway}},\
  }\bibfield  {title} {\emph {\enquote {\bibinfo {title} {Engineering a
  Flux-Dependent Mobility Edge in Disordered Zigzag Chains},}\ }}\href
  {\doibase 10.1103/PhysRevX.8.031045} {\bibfield  {journal} {\bibinfo
  {journal} {Phys. Rev. X}\ }\textbf {\bibinfo {volume} {8}},\ \bibinfo {pages}
  {031045} (\bibinfo {year} {2018})}\BibitemShut {NoStop}%
\bibitem [{\citenamefont {Hsu}\ and\ \citenamefont {Chen}(2020)}]{Hsu20}%
  \BibitemOpen
  \bibfield  {author} {\bibinfo {author} {\bibfnamefont {H.-C.}\ \bibnamefont
  {Hsu}}\ and\ \bibinfo {author} {\bibfnamefont {T.-W.}\ \bibnamefont {Chen}},\
  }\bibfield  {title} {\emph {\enquote {\bibinfo {title} {Topological Anderson
  insulating phases in the long-range Su-Schrieffer-Heeger model},}\ }}\href
  {\doibase 10.1103/PhysRevB.102.205425} {\bibfield  {journal} {\bibinfo
  {journal} {Phys. Rev. B}\ }\textbf {\bibinfo {volume} {102}},\ \bibinfo
  {pages} {205425} (\bibinfo {year} {2020})}\BibitemShut {NoStop}%
\bibitem [{\citenamefont {Ghosh}\ and\ \citenamefont {Nag}(2022)}]{Ghosh22NH}%
  \BibitemOpen
  \bibfield  {author} {\bibinfo {author} {\bibfnamefont {A.~K.}\ \bibnamefont
  {Ghosh}}\ and\ \bibinfo {author} {\bibfnamefont {T.}~\bibnamefont {Nag}},\
  }\bibfield  {title} {\emph {\enquote {\bibinfo {title} {Non-Hermitian
  higher-order topological superconductors in two dimensions: Statics and
  dynamics},}\ }}\href {\doibase 10.1103/PhysRevB.106.L140303} {\bibfield
  {journal} {\bibinfo  {journal} {Phys. Rev. B}\ }\textbf {\bibinfo {volume}
  {106}},\ \bibinfo {pages} {L140303} (\bibinfo {year} {2022})}\BibitemShut
  {NoStop}%
\bibitem [{\citenamefont {Liu}\ \emph {et~al.}(2021)\citenamefont {Liu},
  \citenamefont {Zhou}, \citenamefont {Wu}, \citenamefont {Zhang},\ and\
  \citenamefont {Jiang}}]{Liu21}%
  \BibitemOpen
  \bibfield  {author} {\bibinfo {author} {\bibfnamefont {H.}~\bibnamefont
  {Liu}}, \bibinfo {author} {\bibfnamefont {J.-K.}\ \bibnamefont {Zhou}},
  \bibinfo {author} {\bibfnamefont {B.-L.}\ \bibnamefont {Wu}}, \bibinfo
  {author} {\bibfnamefont {Z.-Q.}\ \bibnamefont {Zhang}}, \ and\ \bibinfo
  {author} {\bibfnamefont {H.}~\bibnamefont {Jiang}},\ }\bibfield  {title}
  {\emph {\enquote {\bibinfo {title} {Real-space topological invariant and
  higher-order topological Anderson insulator in two-dimensional non-Hermitian
  systems},}\ }}\href {\doibase 10.1103/PhysRevB.103.224203} {\bibfield
  {journal} {\bibinfo  {journal} {Phys. Rev. B}\ }\textbf {\bibinfo {volume}
  {103}},\ \bibinfo {pages} {224203} (\bibinfo {year} {2021})}\BibitemShut
  {NoStop}%
\bibitem [{\citenamefont {Liu}\ \emph {et~al.}(2023)\citenamefont {Liu},
  \citenamefont {Lu}, \citenamefont {Zhang},\ and\ \citenamefont
  {Jiang}}]{Liu23}%
  \BibitemOpen
  \bibfield  {author} {\bibinfo {author} {\bibfnamefont {H.}~\bibnamefont
  {Liu}}, \bibinfo {author} {\bibfnamefont {M.}~\bibnamefont {Lu}}, \bibinfo
  {author} {\bibfnamefont {Z.-Q.}\ \bibnamefont {Zhang}}, \ and\ \bibinfo
  {author} {\bibfnamefont {H.}~\bibnamefont {Jiang}},\ }\bibfield  {title}
  {\emph {\enquote {\bibinfo {title} {Modified generalized Brillouin zone
  theory with on-site disorder},}\ }}\href {\doibase
  10.1103/PhysRevB.107.144204} {\bibfield  {journal} {\bibinfo  {journal}
  {Phys. Rev. B}\ }\textbf {\bibinfo {volume} {107}},\ \bibinfo {pages}
  {144204} (\bibinfo {year} {2023})}\BibitemShut {NoStop}%
\bibitem [{\citenamefont {Rangi}\ \emph {et~al.}(2024)\citenamefont {Rangi},
  \citenamefont {Tam},\ and\ \citenamefont {Moreno}}]{RangiPRB2024}%
  \BibitemOpen
  \bibfield  {author} {\bibinfo {author} {\bibfnamefont {C.}~\bibnamefont
  {Rangi}}, \bibinfo {author} {\bibfnamefont {K.-M.}\ \bibnamefont {Tam}}, \
  and\ \bibinfo {author} {\bibfnamefont {J.}~\bibnamefont {Moreno}},\
  }\bibfield  {title} {\emph {\enquote {\bibinfo {title} {Engineering a
  non-Hermitian second-order topological insulator state in quasicrystals},}\
  }}\href {\doibase 10.1103/PhysRevB.109.064203} {\bibfield  {journal}
  {\bibinfo  {journal} {Phys. Rev. B}\ }\textbf {\bibinfo {volume} {109}},\
  \bibinfo {pages} {064203} (\bibinfo {year} {2024})}\BibitemShut {NoStop}%
\bibitem [{\citenamefont {Ji}\ \emph {et~al.}(2024)\citenamefont {Ji},
  \citenamefont {Ding}, \citenamefont {Chen},\ and\ \citenamefont
  {Yang}}]{JiPRB2024}%
  \BibitemOpen
  \bibfield  {author} {\bibinfo {author} {\bibfnamefont {X.}~\bibnamefont
  {Ji}}, \bibinfo {author} {\bibfnamefont {W.}~\bibnamefont {Ding}}, \bibinfo
  {author} {\bibfnamefont {Y.}~\bibnamefont {Chen}}, \ and\ \bibinfo {author}
  {\bibfnamefont {X.}~\bibnamefont {Yang}},\ }\bibfield  {title} {\emph
  {\enquote {\bibinfo {title} {Non-Hermitian second-order topological
  superconductors},}\ }}\href {\doibase 10.1103/PhysRevB.109.125420} {\bibfield
   {journal} {\bibinfo  {journal} {Phys. Rev. B}\ }\textbf {\bibinfo {volume}
  {109}},\ \bibinfo {pages} {125420} (\bibinfo {year} {2024})}\BibitemShut
  {NoStop}%
\bibitem [{\citenamefont {Chiu}\ \emph {et~al.}(2016)\citenamefont {Chiu},
  \citenamefont {Teo}, \citenamefont {Schnyder},\ and\ \citenamefont
  {Ryu}}]{Shinsei16}%
  \BibitemOpen
  \bibfield  {author} {\bibinfo {author} {\bibfnamefont {C.-K.}\ \bibnamefont
  {Chiu}}, \bibinfo {author} {\bibfnamefont {J.~C.~Y.}\ \bibnamefont {Teo}},
  \bibinfo {author} {\bibfnamefont {A.~P.}\ \bibnamefont {Schnyder}}, \ and\
  \bibinfo {author} {\bibfnamefont {S.}~\bibnamefont {Ryu}},\ }\bibfield
  {title} {\emph {\enquote {\bibinfo {title} {Classification of topological
  quantum matter with symmetries},}\ }}\href {\doibase
  10.1103/RevModPhys.88.035005} {\bibfield  {journal} {\bibinfo  {journal}
  {Rev. Mod. Phys.}\ }\textbf {\bibinfo {volume} {88}},\ \bibinfo {pages}
  {035005} (\bibinfo {year} {2016})}\BibitemShut {NoStop}%
\bibitem [{\citenamefont {Lin}\ \emph {et~al.}(2021)\citenamefont {Lin},
  \citenamefont {Ke},\ and\ \citenamefont {Lee}}]{Lin21}%
  \BibitemOpen
  \bibfield  {author} {\bibinfo {author} {\bibfnamefont {L.}~\bibnamefont
  {Lin}}, \bibinfo {author} {\bibfnamefont {Y.}~\bibnamefont {Ke}}, \ and\
  \bibinfo {author} {\bibfnamefont {C.}~\bibnamefont {Lee}},\ }\bibfield
  {title} {\emph {\enquote {\bibinfo {title} {Real-space representation of the
  winding number for a one-dimensional chiral-symmetric topological
  insulator},}\ }}\href {\doibase 10.1103/PhysRevB.103.224208} {\bibfield
  {journal} {\bibinfo  {journal} {Phys. Rev. B}\ }\textbf {\bibinfo {volume}
  {103}},\ \bibinfo {pages} {224208} (\bibinfo {year} {2021})}\BibitemShut
  {NoStop}%
\end{thebibliography}%

\end{document}